\documentclass[journal]{IEEEtran}

\ifCLASSINFOpdf
\else
\fi
\usepackage{algpseudocode}
\usepackage{algorithm}
\usepackage[dvips]{graphicx}
\usepackage{latexsym}
\usepackage{amssymb}
\usepackage{amsmath}
\usepackage{multirow}
\usepackage{xcolor}
\usepackage{lipsum}
\usepackage{multicol}
\usepackage{enumerate}
\usepackage{algorithm}
\usepackage{algorithmicx}
\usepackage{algpseudocode}
\usepackage{amsmath}
\usepackage{graphicx}
\usepackage{subfig}
\usepackage[utf8]{inputenc}
\usepackage{float}
\usepackage{bm}
\begin{document}
%
\title{Interference Exploitation Precoding for Multi-Level Modulations: Closed-Form Solutions}
%
%
%

\author{Ang Li,~\IEEEmembership{Member,~IEEE}, Christos Masouros,~\IEEEmembership{Senior Member,~IEEE}, Yonghui Li,~\IEEEmembership{Senior Member,~IEEE},\\Branka Vucetic~\IEEEmembership{Fellow,~IEEE}, and A. Lee Swindlehurst,~\IEEEmembership{Fellow,~IEEE}

\thanks{Manuscript received TIME; revised TIME.}
\thanks{A. Li was with the Department of Electronic and Electrical Engineering, University College London, Torrington Place, London, WC1E 7JE, UK, and is now with the School of Electrical and Information Engineering, University of Sydney, NSW 2006, Australia. (email: ang.li2@sydney.edu.au)}
\thanks{C. Masouros is with the Department of Electronic and Electrical Engineering, University College London, Torrington Place, London, WC1E 7JE, UK. (email: c.masouros@ucl.ac.uk)}
\thanks{Y. Li and B. Vucetic are with the School of Electrical and Information Engineering, University of Sydney, Sydney, NSW 2006, Australia. (e-mail: \{yonghui.li, branka.vucetic\}@sydney.edu.au)}
\thanks{A. L. Swindlehurst is with the Department of Electrical Engineering and Computer Science, Henry Samueli School of Engineering, University of California, Irvine, CA 92697, USA. (e-mail: swindle@uci.edu)}
\thanks{This work was supported by the Royal Academy of Engineering, UK, the Engineering and Physical Sciences Research Council (EPSRC) project EP/M014150/1 and the China Scholarship Council (CSC).}
}



\maketitle

\begin{abstract}
In this paper, we study closed-form interference-exploitation precoding for multi-level modulations in the downlink of multi-user multiple-input single-output (MU-MISO) systems. We consider two distinct cases: first, for the case where the number of served users is not larger than the number of transmit antennas at the base station (BS), we mathematically derive the optimal precoding structure based on the Karush-Kuhn-Tucker (KKT) conditions. By formulating the dual problem, the precoding problem for multi-level modulations is transformed into a pre-scaling operation using quadratic programming (QP) optimization. We further consider the case where the number of served users is larger than the number of transmit antennas at the BS. By employing the pseudo inverse, we show that the optimal solution of the pre-scaling vector is equivalent to a linear combination of the right singular vectors corresponding to zero singular values, and derive the equivalent QP formulation. We also present the condition under which multiplexing more streams than the number of transmit antennas is achievable. For both considered scenarios, we propose a modified iterative algorithm to obtain the optimal precoding matrix, as well as a sub-optimal closed-form precoder. Numerical results validate our derivations on the optimal precoding structures for multi-level modulations, and demonstrate the superiority of interference-exploitation precoding for both scenarios.
\end{abstract}


%
\IEEEpeerreviewmaketitle

\section{Introduction}
%
%
%
%
\IEEEPARstart{P}{RECODING} has been widely studied in multi-antenna wireless communication systems to simultaneously support data transmission to multiple users \cite{r1}. When the channel state information (CSI) is known at the transmitter side, dirty paper coding (DPC) that subtracts the interference prior to transmission achieves the channel capacity \cite{r2}. Despite its promising performance, DPC is generally difficult to implement in practical wireless systems, due to its impractical assumption of an infinite source alphabet and prohibitive complexity. Therefore, sub-optimal approximations of DPC in the form of Tomlinson-Harashima precoding (THP) and vector perturbation (VP) precoding have been proposed in \cite{r3} and \cite{r4}, respectively. While offering near-optimal performance, both THP and VP are still non-linear precoding methods and include a sphere-search process, which makes their complexity still unfavorable, especially when the number of data streams is large. Accordingly, low-complexity linear precoding methods such as zero-forcing (ZF) \cite{r5} and regularized ZF (RZF) \cite{r6} have become popular. On the other hand, downlink precoding based on optimization has also received increasing research attention \cite{r7}\nocite{r8}\nocite{r9}\nocite{r10}\nocite{r11}\nocite{r12}-\cite{r13}. Among optimization-based precoding methods, the two most well-known designs are referred to as signal-to-noise-plus-interference ratio (SINR) balancing \cite{r7}-\cite{r9} and power minimization \cite{r10}-\cite{r12}, where SINR balancing aims to maximize the minimum received SINR subject to a total transmit power constraint \cite{r7}, \cite{r8} or a per-antenna power constraint \cite{r9}, and power minimization targets minimizing the power consumption at the transmitter side while guaranteeing a minimum SINR at each receiver \cite{r11}. 

For both the closed-form precoding schemes and the optimization-based precoding approaches described above, the CSI at the base station (BS) is exploited to design the precoding strategy that eliminates, avoids or limits interference. The above approaches ignore the fact that the information in the transmitted data symbols themselves can also be exploited in the downlink precoding design on a symbol-by-symbol basis for further performance improvements. With information about the data symbols and their corresponding constellations, the instantaneous interference can be divided into constructive interference (CI) and destructive interference \cite{r30}. More specifically, CI is defined as interference that pushes the received signals away from the detection thresholds \cite{r14}, \cite{r15}, which provides further benefits for signal detection. A modified ZF precoding method was proposed in \cite{r16} to exploit the constructive part of the interference while eliminating the destructive part. A more advanced two-stage interference exploitation precoding was proposed in \cite{r17}, where the phase of the destructive interference was controlled and further rotated such that the destructive interference becomes constructive. Optimization-based interference-exploitation precoding for PSK modulations has also been proposed in \cite{r31} in the context of vector perturbation precoding, where CI in the form of symbol scaling is proposed. In \cite{r18}\nocite{r19}-\cite{r20}, CI precoding based on the phase-rotation metric is studied, where it is shown that a relaxed non-strict phase rotation metric is more advantageous compared to the strict phase rotation in \cite{r16}, \cite{r17}. For multi-level modulations such as QAM, CI can be exploited for the outer constellation points, although all the interference for the inner constellation points is considered to be destructive, as discussed in \cite{r26}\nocite{r48}-\cite{r21} where a symbol-scaling metric is introduced. Due to the above benefits, CI has been extended to the area of low-resolution digital-to-analog converters (DACs) with PSK signaling in \cite{r38}, as well as quantized constant envelope precoding with PSK and QAM signaling in \cite{r49}. More recently, it has been revealed in \cite{r45} that there exists an optimal structure for the CI precoding for PSK modulations. Nevertheless, it is still unclear whether a similar result exists for multi-level modulations such as QAM, since CI precoding for PSK modulations is based on the phase-rotation metric, while the symbol-scaling metric has to be employed for QAM constellations.

In this paper, we study closed-form interference exploitation precoding for multi-level modulations, where QAM modulation is considered as a representative example. Due to the fact that the conventional phase-rotation CI formulation is not applicable to QAM constellations, the more general symbol-scaling metric is employed. We reveal the geometric connection between the phase-rotation and symbol-scaling metrics in the CI formulation, based on which we propose the optimization problem that maximizes the CI effect for the outer constellation symbols while constraining the inner constellation symbols for multi-level modulations. We first study the case where the number of users simultaneously served by the BS is not larger than the number of BS transmit antennas. Using the Lagrangian and KKT conditions, we analyze the formulated problem and mathematically derive the structure of the optimal precoding matrix, which leads to an equivalent simplified optimization problem. By further formulating the dual problem of this equivalent optimization, we show that, similar to the case of PSK modulations, interference-exploitation precoding for multi-level modulations is equivalent to a quadratic programming (QP) optimization, and the optimal precoding matrix can be expressed as a function of the dual variables in closed form. 

We further extend our analysis to the case where the number of users simultaneously served by the BS is larger than the number of BS transmit antennas, in which case conventional precoding becomes infeasible and the exact inverse included in the above analysis becomes inapplicable. In this scenario, we show that interference-exploitation precoding may still be feasible. To this end, the more generic pseudo inverse of the channel matrix is employed instead, and we derive the optimal structure of the precoding matrix. Due to the inclusion of the pseudo inverse, an additional constraint is further introduced in the equivalent optimization. Built upon this, the scaling vector for the constellation symbols is shown to be the non-zero solution of a linear equation set, which is equivalent to a linear combination of the singular vectors corresponding to the zero singular values of the coefficient matrix. Accordingly, the optimization can be transformed into an optimization on the weights for each singular vector, which is further shown to be equivalent to a QP optimization as well. Based on the equivalent QP formulation, we discuss the condition under which multiplexing more streams than the number of transmit antennas is possible with interference exploitation precoding. 

For both of the scenarios considered above, we also present a generic iterative algorithm to efficiently obtain the optimal precoding matrix for multi-level modulations, where a closed-form update is included in each iteration. Based on the above transformation and algorithm, we further develop a sub-optimal closed-form non-iterative CI precoder. Our analysis for multi-level modulations in this paper complements the study on closed-form symbol-level interference-exploitation precoding in \cite{r45}, which is not applicable to multi-level modulations. Simulation results validate our mathematical derivations and the optimality of the proposed algorithm. Moreover, the superiority of interference-exploitation precoding over conventional precoding methods for multi-level modulations is also revealed, especially for the case where the BS simultaneously serves a larger number of users than it has the number of transmit antennas.

We summarize the contributions of this paper below:
\begin{enumerate}

\item We present a geometric connection between symbol-scaling and phase-rotation metrics for interference-exploitation precoding, based on which we construct the optimization that maximizes the CI effect of the outer constellation symbols while maintaining the performance of the inner constellation symbols for multi-level modulations. 

\item We perform mathematical analysis on interference-exploitation precoding for multi-level modulations. We show that CI precoding for multi-level modulations can ultimately be simplified into a QP optimization as well. Compared to CI precoding for PSK modulations where the optimization is over a simplex, it is shown that only part of the dual variables need to be constrained as non-negative in the QP formulation for multi-level modulations.

\item We further extend our analysis on CI to the case where the number of served users is larger than the number of transmit antennas at the BS. Our transformations show that the optimization for CI precoding in such scenarios is similar to the conventional case where the number of users is smaller than or equal to the number of antennas at the BS, also resulting in a QP optimization. We also present the condition under which multiplexing more streams than the number of transmit antennas based on CI is achievable.

\item We propose an iterative algorithm that is able to obtain the optimal solution of a generic QP optimization problem subject to specific constraints within only a few iterations. Based on this algorithm, the optimal precoding matrix can be efficiently obtained, for both scenarios considered in this paper. A sub-optimal closed-form non-iterative precoder is also presented.

\end{enumerate}

The remainder of this paper is organized as follows: Section II introduces the system model and illustrates the connection between the two CI metrics. Section III includes the CI-based optimization problems for multi-level modulations when the number of users is smaller than or equal to the number of BS transmit antennas, and the extension to the scenario when the number of users is larger than the number of BS transmit antennas is studied in Section IV. The modified iterative algorithm and sub-optimal closed-form precoder are presented in Section V. Numerical results are provided in Section VI, and Section VII concludes the paper.

{\it Notation}: $a$, $\bf a$, and $\bf A$ denote scalar, column vector and matrix, respectively. ${( \cdot )^*}$, ${( \cdot )^T}$, ${( \cdot )^H}$, ${( \cdot )^{-1}}$, ${( \cdot )^+}$ and ${\rm rank} \left\{ \cdot \right\}$ denote conjugate, transposition, conjugate transposition, inverse, pseudo inverse, and rank of a matrix, respectively. $diag \left(  \cdot  \right)$ is the transformation of a column vector into a diagonal matrix, and ${\rm{vec}}\left(  \cdot  \right)$ denotes the vectorization operation. ${\bf{A}}({k,i})$ denotes the entry in the $k$-row and $i$-th column of $\bf A$. $\left|  \cdot  \right|$ denotes the absolute value of a real number or the modulus of a complex number, and $\left\|  \cdot  \right\|_2$ denotes the $\ell_2$-norm. ${{\cal C}^{n \times n}}$ and ${{\cal R}^{n \times n}}$ represent the sets of $n\times n$ complex- and real-valued matrices, respectively. $\Re \left\{ \cdot \right\}$ and $\Im \left\{ \cdot \right\}$ respectively denote the real and imaginary part of a complex scalar, vector or matrix. $card\left\{  \cdot  \right\}$ denotes the cardinality of a set, and $\otimes$ represents the Kronecker product. $j$ denotes the imaginary unit, ${\bf I}_K$ denotes the $K \times K$ identity matrix, and ${\bf e}_i$ represents the $i$-th column of the identity matrix.

\section{System Model and Constructive Interference}
\subsection{System Model}
We study a downlink MU-MISO system, where the BS with $N_t$ transmit antennas is simultaneously communicating with $K$ single-antenna users in the same time-frequency resource. We separately consider the scenarios of both $K \le N_t$ and $K>N_t$. We focus on the downlink precoding designs, and perfect CSI is assumed throughout the paper. The data symbol vector is assumed to be from a normalized multi-level modulation constellation \cite{r18}, denoted as ${\bf s} \in {\cal C}^{K \times 1}$, and the received signal at the $k$-th user can then be expressed as
\begin{equation}
r_k= {\bf h}_k^T{\bf Ws} + n_k,
\label{eq_1}
\end{equation}
where ${\bf h}_k \in {\cal C}^{N_t \times 1}$ denotes the flat-fading Rayleigh channel vector from user $k$ to the BS with each entry following a standard complex Gaussian distribution, ${\bf W} \in {\cal C}^{N_t \times K}$ is the precoding matrix, and $n_k$ is the additive Gaussian noise at the receiver with zero mean and variance $\sigma^2$. 

\begin{figure}[!b]
\centering
\includegraphics[scale=0.35]{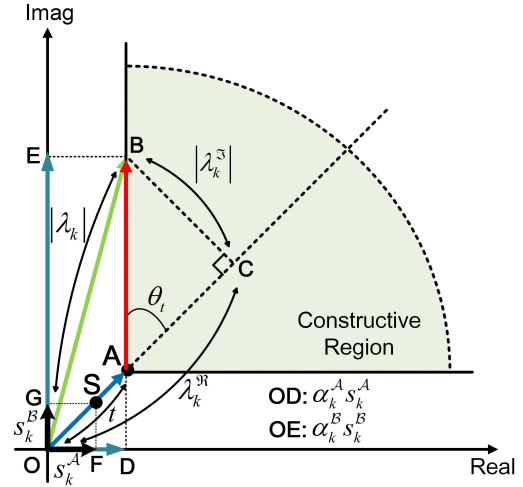}
\caption{Symbol-scaling and phase-rotation metric for QPSK constellation}
\end{figure}

\subsection{Connection between Two CI Metrics for PSK Modulation}
In this section, we illustrate the connection between symbol-scaling and phase-rotation metrics for CI precoding based on Fig. 1, where we employ QPSK (4QAM) as an example.

{\it Phase Rotation Metric:} As discussed in \cite{r45}, we denote $\vec{OS}=s_k$ and $\vec{OA}=t \cdot s_k$, where $t=\frac{{|{\vec {OA}}|}}{{|{\vec {OS}}|}}$ is the objective to be optimized. We further denote $\vec{OB}$ as the received signal for user $k$ excluding noise, which leads to
\begin{equation}
\vec{OB}={\bf h}_k^T{\bf Ws}=\lambda_k s_k,
\label{eq_2}
\end{equation}
where $\lambda_k$ is a complex scalar that represents the effect of interference on the data symbol for user $k$. For ${\cal M}$-PSK constellations, the CI constraint is then constructed as \cite{r45}
\begin{equation}
\left( {\lambda _k^\Re  - t} \right)\tan {\theta _t} \ge \left| {\lambda _k^\Im } \right|,
\label{eq_3}
\end{equation}
where $\lambda _k^\Re=\Re \left( {{\lambda _k}} \right)$, $\lambda _k^\Im=\Im \left( {{\lambda _k}} \right)$, and ${\theta _t} = \frac{\pi }{\cal M}$ for $\cal M$-PSK constellations. Accordingly, the optimization problem that maxmizes the distance of the constructive region to the detection thresholds subject to the total available transmit power $p_0$ based on the phase-rotation CI metric can be formulated as \cite{r45}
\begin{equation}
\begin{aligned}
&\mathcal{P}_1: {\kern 3pt} \mathop {\max }\limits_{{{\bf{W}}}, {\kern 1pt} t} {\kern 3pt} t \\
&{\kern 0pt} s. t. {\kern 10pt} {{\bf{h}}_k^T}{\bf Ws} = {\lambda _k}{s_k}, {\kern 3pt} \forall k \in {\cal K} \\
&{\kern 22pt} \left( {\lambda _k^\Re  - t} \right)\tan {\theta _t} \ge \left| {\lambda _k^\Im } \right|, {\kern 3pt} \forall k \in {\cal K}\\
&{\kern 22pt} \left\| {{\bf{Ws}}} \right\|_2^2 \le {p_0}
\label{eq_4}
\end{aligned}
\end{equation}
where ${\cal K}=\left\{ {1,2,\cdots,K} \right\}$. We have enforced a symbol-level power constraint on the precoder, since the exploitation of CI is dependent on the data symbol $\bf s$, which will also be shown mathematically in the following. 

{\it Symbol Scaling Metric:} Following the coordinate transformation approach in \cite{r38}, we first decompose the data symbol along the detection thresholds for each user $k$, expressed as
\begin{equation}
\vec{OS}=\vec{OF}+\vec{OG} \Rightarrow s_k=s_k^{\cal A} + s_k^{\cal B},
\label{eq_5}
\end{equation}
where $s_k^{\cal A}$ and $s_k^{\cal B}$ are the bases that are parallel to the detection thresholds for each specific constellation symbol, as shown in Fig. 1. We refer the interested readers to \cite{r38} for a detailed derivation of the expressions for $s_k^{\cal A}$ and $s_k^{\cal B}$ for generic PSK constellations. Specifically for QPSK modulation considered in Fig. 1 as well as QAM modulations in the following part of the paper, we can obtain 
\begin{equation}
s_k^{\cal A}=\Re \left\{ {s_k}\right\} = s_k^\Re, {\kern 3pt} s_k^{\cal B}=j \cdot \Im \left\{ {s_k}\right\} = j \cdot s_k^\Im.
\label{eq_6}
\end{equation}
Following a similar approach to \eqref{eq_5}, we also decompose the noiseless received signal for each user $k$ along the same detection thresholds, and further introduce two real scalars $\alpha_k^{\cal A}$ and $\alpha_k^{\cal B}$ for $s_k^{\cal A}$ and $s_k^{\cal B}$, respectively, which leads to
\begin{equation}
\vec{OB}=\vec{OD}+\vec{OE} \Rightarrow {{\bf{h}}_k^T}{\bf Ws} = \alpha_k^{\cal A}s_k^{\cal A} + \alpha_k^{\cal B}s_k^{\cal B}.
\label{eq_7}
\end{equation}
It is then observed that the values of these two scalars directly indicate the effect of the CI. Subsequently, the corresponding optimization based on the symbol-scaling metric can be constructed as
\begin{equation}
\begin{aligned}
&\mathcal{P}_2: {\kern 3pt} \mathop {\max }\limits_{\bf{W}} \mathop {\min }\limits_{k, {\kern 1pt} {\cal U}} {\kern 3pt} \alpha_k^{\cal U} \\
&{\kern 0pt} s. t. {\kern 10pt} {{\bf{h}}_k^T}{\bf Ws} = \alpha_k^{\cal A}s_k^{\cal A} + \alpha_k^{\cal B}s_k^{\cal B}, {\kern 3pt} \forall k \in {\cal K} \\
&{\kern 22pt} \left\| {{\bf{Ws}}} \right\|_2^2 \le {p_0} \\
&{\kern 25pt} {\cal U} \in \left\{ {{\cal A}, {\cal B}} \right\}
\label{eq_8}
\end{aligned}
\end{equation}
Both of the above optimization problems are convex and can be directly solved with convex optimization tools. Subsequently, based on Fig. 1 and the formulation of the above two optimizations, an important geometrical observation is given, which demonstrates the connection between the symbol-scaling and phase-rotation metric.

{\bf Observation 1:} Since the noiseless received signal is located on the boundary of its constructive region, the relationship between the minimum value of $\left({\alpha_k^{\cal U}}\right)^*$ in ${\cal P}_2$ and the optimal value of $t^*$ in ${\cal P}_1$ is expressed as
\begin{equation}
t^* = 2\left(\alpha _k^{\cal U}\right)^*\left| {\left( {s_k^{\cal U}} \right)^*} \right|\cos \frac{\pi }{\cal M},
\label{eq_9}
\end{equation}
where without loss of generality we have assumed user $k$ has the minimum value of $\alpha$. Eq. \eqref{eq_9} is derived by considering the isosceles triangle `DOA', where we can obtain
\begin{equation}
|\vec{OA}|=2 |\vec{DO}| \cos \angle_{DOA}.
\label{eq_10}
\end{equation}
Based on the fact that $|\vec{OA}|=t^*$, $|\vec{DO}|=|\vec{DA}|=\left( {\alpha_k^{\cal U}} \right)^* \left| {\left( {s_k^{\cal U}} \right)^*} \right|$, and $\angle_{DOA}=\frac{\pi}{\cal M}$, \eqref{eq_10} leads to the expression for $t^*$ in \eqref{eq_9}. {\kern 190pt} $\blacksquare$

It's worth noting that while the above discussion only focuses on QPSK constellations, \eqref{eq_9} is in fact generic to any ${\cal M}$-PSK modulation for the connection between the two CI metrics, and the only difference lies in the expression for $s_k^{\cal U}$. In the following section, the symbol-scaling CI metric is employed in the derivation of the optimal precoding matrix for multi-level modulations. 

\section{CI Precoding for the Case of $K \le N_t$}
In this section, we focus on the common case where $K \le N_t$, and we consider 16QAM modulation as an example of multi-level modulations. For other multi-level constellations, the problem formulation and the corresponding analysis for the symbol-scaling metric readily follows our derivations in this section in a similar way. 

\begin{figure}[!t]
\centering
\includegraphics[scale=0.6]{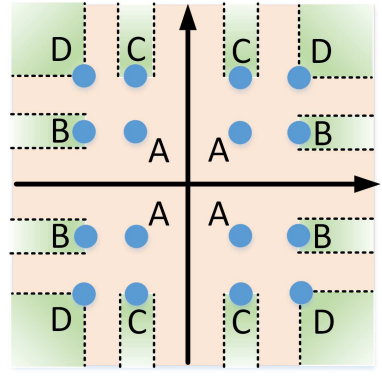}
\caption{Constellation point categorization for 16QAM}
\end{figure}

For a generic QAM constellation, we employ the symbol-scaling metric for CI precoding since there does not exist a generic expression for the phase-rotation CI metric for QAM modulations, as shown in Fig. 2 where a 16QAM constellation is depicted as the example. The symbol-scaling metric in \eqref{eq_7} can be further expressed in vector form as
\begin{equation}
{{\bf{h}}_k^T}{\bf Ws}={{\bf \Omega} _k^T}{{\bf s}_k},
\label{eq_11}
\end{equation}
where we have introduced two column vectors 
\begin{equation}
{{{\bf \Omega}_k}}=\left[ {\alpha_k^{\cal A}, \alpha_k^{\cal B}} \right]^T, {\kern 3pt} {{\bf{s}}_k}=\left[ {s_k^{\cal A}, s_k^{\cal B}}\right]^T.
\label{eq_12}
\end{equation} 
For QAM constellations, $s_k^{\cal A}$ and $s_k^{\cal B}$ are also given by \eqref{eq_6}. In this work, we consider the interference on the inner constellation points as only destructive, since the interference is less likely to be beneficial for these points. To be more specific, in Fig.2 CI exists for the real part of the constellation point type `B' and imaginary part of type `C', while both the real and imaginary part of the constellation point type `D' can be exploited. Accordingly, we propose to construct the optimization problem that maximizes the CI effect for the outer constellation points while maintaining the performance for the inner constellation points, given by
\begin{equation}
\begin{aligned}
&\mathcal{P}_3: {\kern 3pt} \mathop {\max }\limits_{{{\bf{W}}}, {\kern 1pt} t} {\kern 3pt} t \\
&{\kern 0pt} s. t. {\kern 10pt} {{\bf{h}}_k^T}{\bf Ws} = {{\bf \Omega} _k^T}{{\bf{s}}_k}, {\kern 3pt} \forall k \in {\cal K} \\
&{\kern 24pt} t \le \alpha_m^{\cal O}, {\kern 3pt} \forall \alpha_m^{\cal O} \in {\cal O}\\
&{\kern 24pt} t = \alpha_n^{\cal I}, {\kern 3pt} \forall \alpha_n^{\cal I} \in {\cal I}\\
&{\kern 22pt} \left\| {{\bf{Ws}}} \right\|_2^2 \le {p_0}
\label{eq_13}
\end{aligned}
\end{equation}
where the set ${\cal O}$ consists of the real scalars corresponding to the real or imaginary part of the outer constellation points that can be scaled, and ${\cal I}$ consists of the real scalars corresponding to the real or imaginary part of the constellation points that cannot exploit CI. Accordingly, we obtain 
\begin{equation}
{\cal O} \cup {\cal I} = \left\{ {\alpha _1^{\cal A} ,\alpha _1^{\cal B} ,\alpha _2^{\cal A} ,\alpha _2^{\cal B} , \cdots ,\alpha _K^{\cal A} ,\alpha _K^{\cal B} } \right\},
\label{eq_14}
\end{equation}
and 
\begin{equation}
card\left\{ {\cal O} \right\} + card\left\{ {\cal I} \right\} = 2K.
\label{eq_15}
\end{equation}

${\cal P}_3$ is a second-order-cone programming (SOCP) problem, which can be solved via convex optimization tools such as CVX. Specifically, the optimization objective $t$ is equal to the value of $\alpha_k^{\cal I}$ in the above optimization, which can also be viewed as a scaling factor for the constellation. Moreover, if we further constrain $t = \alpha_m^{\cal O}$ instead of $t \le \alpha_m^{\cal O}$ in the above optimization, the solution of the above optimization problem will become a ZF precoder.

Before we present the subsequent analysis, we first transform the power constraint included in the above optimization problem, which greatly simplifies the subsequent derivations. To be specific, we decompose the precoded signals $\bf Ws$ into
\begin{equation}
{\bf Ws}=\sum\limits_{i = 1}^K {{{\bf{w}}_i}{s_i}},
\label{eq_16}
\end{equation}
and similar to the case of PSK \cite{r45}, we observe that the distribution of the power among each ${\bf w}_i s_i$ does not affect the solution of the above optimization problem, as ${\bf Ws}$ can be viewed as a single vector for both constraints that include $\bf W$ in ${\cal P}_3$. Therefore, without loss of generality and to be consistent with our problem formulation for PSK modulation in \cite{r45}, we assume that the norm of each term ${\bf w}_i s_i$ is identical, and we obtain
\begin{equation}
\begin{aligned}
&\left\| {{\bf{Ws}}} \right\|_2^2 = {K^2} \left\| {{{\bf{w}}_i}{s_i}} \right\|_2^2 = {K^2} s_i^*{\bf{w}}_i^H{{\bf{w}}_i}{s_i}, \\
&{\kern 15pt}\sum\limits_{i = 1}^K {s_i^*{\bf{w}}_i^H{{\bf{w}}_i}{s_i}}  = K  s_i^*{\bf{w}}_i^H{{\bf{w}}_i}{s_i},
\label{eq_17}
\end{aligned}
\end{equation}
which further leads to the equivalent power constraint as
\begin{equation}
\sum\limits_{i = 1}^K {s_i^*{\bf{w}}_i^H{{\bf{w}}_i}{s_i}} \le \frac{p_0}{K}.
\label{eq_18}
\end{equation}
We then rewrite the above optimization problem ${\cal P}_3$ in standard minimization form as
\begin{equation}
\begin{aligned}
&\mathcal{P}_4: {\kern 3pt} \mathop {\min }\limits_{{{\bf{W}}}, {\kern 1pt} t} {\kern 3pt} -t \\
&{\kern 0pt} s. t. {\kern 10pt} {{\bf{h}}_k^T}\sum\limits_{i = 1}^K {{{\bf{w}}_i}{s_i}} - {{\bf \Omega} _k^T}{{\bf{s}}_k}=0, {\kern 3pt} \forall k \in {\cal K} \\
&{\kern 24pt} t - \alpha_m^{\cal O} \le 0, {\kern 3pt} \forall \alpha_m^{\cal O} \in {\cal O}\\
&{\kern 24pt} t - \alpha_n^{\cal I} = 0 , {\kern 3pt} \forall \alpha_n^{\cal I} \in {\cal I}\\
&{\kern 22pt} \sum\limits_{i = 1}^K {s_i^*{\bf{w}}_i^H{{\bf{w}}_i}{s_i}} \le \frac{p_0}{K}
\label{eq_19}
\end{aligned}
\end{equation}
and we express the Lagrangian of ${\cal P}_4$ as \cite{r27}
\begin{equation}
\begin{aligned}
&{\cal L} \left( {{{\bf{w}}_i},t,{\delta _k},{\mu _m},{\nu _n},{\delta _0}} \right) =  - t \\
&+ \sum\limits_{k = 1}^K {{\delta _k}\left( {{{\bf{h}}_k^T}\sum\limits_{i = 1}^K {{{\bf{w}}_i}{s_i}}  - {{\bf \Omega}_k^T}{{{\bf{s}}}_k}} \right)}  + \sum\limits_{m = 1}^{card\left\{ {\cal O} \right\}} {{\mu _m}\left( {t - \alpha _m^{\cal O}} \right)}  \\
&+ \sum\limits_{n = 1}^{card\left\{ {\cal I} \right\}} {{\nu _n}\left( {t - \alpha _n^{\cal I}} \right)}  + {\delta _0}\left( {\sum\limits_{i = 1}^K {s_i^*{\bf{w}}_i^H{{\bf{w}}_i}{s_i}}  - \frac{{{p_0}}}{K}} \right),
\end{aligned}
\label{eq_20}
\end{equation}
where ${\delta _k}$, ${\mu _m}$, ${\nu _n}$, and ${\delta _0}$ are the introduced dual variables, $\delta_0 \ge 0$ and ${\mu _m} \ge 0$, $\forall m \in \left\{ {1,2,\cdots,card\left\{ {\cal O} \right\}} \right\}$. Each ${\delta _k}$ and $\nu_n$ can be complex since they correspond to the equality constraints. 

Based on the Lagrangian in \eqref{eq_20}, the KKT conditions for optimality can be expressed as
\begin{IEEEeqnarray}{rCl} 
\IEEEyesnumber
\frac{{\partial {\cal L}}}{{\partial t}} =  - 1  + \sum\limits_{m = 1}^{card\left\{ {\cal O} \right\}} {{\mu _m}}  + \sum\limits_{n = 1}^{card\left\{ {\cal I} \right\}} {{\nu _n}}   = 0 {\kern 30pt} \IEEEyessubnumber* \label{eq_21a} \\
\frac{{\partial {\cal L}}}{{\partial {{\bf{w}}_i}}} = \left( {\sum\limits_{k = 1}^K {{\delta _k} \cdot {{\bf{h}}_k^T}} } \right){s_i} + {\delta _0} s_i s_i^* \cdot {\bf{w}}_i^H = {\bf{0}}, {\kern 2pt} \forall i \in {\cal K} {\kern 30pt} \label{eq_21b} \\
{{{\bf{h}}_k^T}\sum\limits_{i = 1}^K {{{\bf{w}}_i}{s_i}}  - {{\bf \Omega}_k^T}{{{\bf{s}}}_k}} = 0, {\kern 2pt} \forall k \in {\cal K}{\kern 30pt} \label{eq_21c} \\
{\mu _m}\left( {t - {\alpha _m^{\cal O}}} \right) = 0, {\kern 2pt} \forall \alpha _m^{\cal O} \in {\cal O} {\kern 30pt} \label{eq_21d}\\
t - {\alpha _n^{\cal I}} = 0, {\kern 2pt} \forall \alpha _n^{\cal I} \in {\cal I} {\kern 30pt} \label{eq_21e}\\
{\delta _0}\left( {\sum\limits_{i = 1}^K {s_i^*{\bf{w}}_i^H{{\bf{w}}_i}{s_i}}  - \frac{p_0}{K}} \right) = 0 {\kern 30pt} \label{eq_21f}
\end{IEEEeqnarray}
Based on \eqref{eq_21b}, it is first observed that $\delta_0 \ne 0$, and with the premise that $\delta_0 \ge 0$ we obtain $\delta_0 >0$, which further means that the power constraint is met with equality when optimality is achieved. Then, we can express ${\bf w}_i^H$ in \eqref{eq_21b} as
\begin{equation}
{\bf{w}}_i^H =  - \frac{{{s_i}}}{{{\delta _0}{s_i}s_i^*}}\left( {\sum\limits_{k = 1}^K {{\delta _k} \cdot {{\bf{h}}_k^T}} } \right) =  - \frac{1}{{s_i^*}}\left( {\sum\limits_{k = 1}^K {\frac{{{\delta _k}}}{{{\delta _0}}} \cdot {{\bf{h}}_k^T}} } \right).
\label{eq_22}
\end{equation}
By introducing an auxiliary variable
\begin{equation}
{\vartheta _k} =  - \frac{{{\delta _k^H}}}{{{\delta _0}}}, {\kern 3pt} \forall k \in {\cal K},
\label{eq_23}
\end{equation}
we can express ${\bf w}_i$ as
\begin{equation}
{{\bf{w}}_i} = \left( {\sum\limits_{k = 1}^K {{\vartheta _k} \cdot {\bf{h}}_k^*} } \right) \frac{1}{{{s_i}}}, {\kern 3pt} \forall i \in {\cal K}.
\label{eq_24}
\end{equation}
The above expression further leads to
\begin{equation}
{{\bf{w}}_i}s_i = \left( {\sum\limits_{k = 1}^K {{\vartheta _k} \cdot {\bf{h}}_k^*} } \right), {\kern 3pt} \forall i \in {\cal K},
\label{eq_25}
\end{equation}
which is constant for any $i$ and consistent with our assumption in \eqref{eq_17}. 

With the obtained expression for each ${{\bf{w}}_i}$, we further express the precoding matrix $\bf W$ as
\begin{equation}
\begin{aligned}
{\bf{W}} &= \left[ {{{\bf{w}}_1},{{\bf{w}}_2}, \cdots ,{{\bf{w}}_K}} \right]\\
&=\left( {\sum\limits_{k = 1}^K {{\vartheta _k} \cdot {\bf{h}}_k^*} } \right)  \left[ {\frac{1}{{{s_1}}},\frac{1}{{{s_2}}}, \cdots ,\frac{1}{{{s_K}}}} \right]\\
&=\left[ {{\bf{h}}_1^*,{\bf{h}}_2^*, \cdots ,{\bf{h}}_K^*} \right] {\left[ {{\vartheta _1},{\vartheta _2}, \cdots ,{\vartheta _K}} \right]^T} \left[ {\frac{1}{{{s_1}}},\frac{1}{{{s_2}}}, \cdots ,\frac{1}{{{s_K}}}} \right] \\
&={\bf H}^H {\bf{\Upsilon}} {\bf \hat s}^T,
\end{aligned}
\label{eq_26}
\end{equation}
where we have introduced two column vectors
\begin{equation}
{\bf{\Upsilon}}={\left[ {{\vartheta _1},{\vartheta _2}, \cdots ,{\vartheta _K}} \right]^T}, {\kern 3pt} {\bf \hat s}=\left[ {\frac{1}{{{s_1}}},\frac{1}{{{s_2}}}, \cdots ,\frac{1}{{{s_K}}}} \right]^T.
\label{eq_27}
\end{equation}
We express \eqref{eq_11} in matrix form as
\begin{equation}
\begin{aligned}
{\bf{HWs}} &= {\left[ {{\bf \Omega} _1^T{{{\bf{s}}}_1},{\bf \Omega} _2^T{{{\bf{s}}}_2}, \cdots, {\bf \Omega}_K^T{{{\bf{s}}}_K}} \right]^T} \\
& = {\bf U} diag \left( {\bf \Omega} \right) {\bf s_E},
\end{aligned}
\label{eq_28}
\end{equation}
where ${\bf \Omega} \in {\cal R}^{2K \times 1}$ and ${\bf s_E} \in {\cal R}^{2K \times 1} $ are expressed as
\begin{equation}
\begin{aligned}
{\bf \Omega}&={\left[ {{\bf \Omega}_1^T, {\bf \Omega} _2^T, \cdots , {\bf  \Omega}_K^T} \right]^T} \\
&= \left[ {\alpha _1^{\cal A} ,\alpha _1^{\cal B} ,\alpha _2^{\cal A} ,\alpha _2^{\cal B} , \cdots ,\alpha _K^{\cal A} ,\alpha _K^{\cal B} } \right]^T\\
&=\left[ {\alpha _1^{E} , \alpha _2^{E} , \cdots ,\alpha _{2K-1}^{E}, \alpha _{2K}^{E} } \right]^T,\\
{\bf{s}_E} &= {\left[ {{\bf{s}}_1^T,{\bf{s}}_2^T, \cdots ,{\bf{s}}_K^T} \right]^T} \\
& = {\left[ {s_1^{\cal A} , s_1^{\cal B} ,s_2^{\cal A} ,  s_2^{\cal B} , \cdots ,s_K^{\cal A} , s_K^{\cal B} } \right]^T}\\
&= {\left[ {s_1^{E} ,s_2^{E} , \cdots ,s_{2K-1}^{E} , s_{2K}^{E} } \right]^T} ,
\end{aligned}
\label{eq_29}
\end{equation}
and the matrix ${\bf U} \in {\cal C}^{K \times 2K}$ is constructed as
\begin{equation}
{\bf{U}} = \left[ {\begin{array}{*{20}{c}}
1&1&0&0& \cdots &0&0\\
0&0&1&1& \ddots & \vdots & \vdots \\
 \vdots & \vdots & \ddots & \ddots & \ddots & \vdots & \vdots \\
 \vdots & \vdots & \ddots & \ddots & \ddots &0&0\\
0&0& \cdots &0&0&1&1
\end{array}} \right] = {\bf I} \otimes \left[ {1,1} \right].
\label{eq_30}
\end{equation}

By substituting the expression for $\bf W$ in \eqref{eq_26} into \eqref{eq_28}, we obtain
\begin{equation}
{\bf{H}}{{\bf{H}}^H} {\bf{\Upsilon}} {{\bf{\hat s}}^T} {\bf s} = {\bf{U}}diag\left( {\bf \Omega} \right){\bf s_E}.
\label{eq_31}
\end{equation}
With the premise that $K \le N_t$ in this section, ${\bf HH}^H$ is invertible, and accordingly we obtain ${\bf{\Upsilon}}$ as 
\begin{equation}
{\bf{\Upsilon}} = \frac{1}{K} \cdot {\left( {{\bf{H}}{{\bf{H}}^H}} \right)^{ - 1}}{\bf{U}}diag\left( {\bf \Omega} \right){\bf s_E},
\label{eq_32}
\end{equation}
which further leads to the expression for the precoding matrix $\bf W$ as
\begin{equation}
{\bf W}=\frac{1}{K} \cdot {{\bf{H}}^H}{\left( {{\bf{H}}{{\bf{H}}^H}} \right)^{ - 1}}{\bf{U}}diag\left( {\bf \Omega} \right){\bf s_E}{{\bf{\hat s}}^T}.
\label{eq_33}
\end{equation}
We then substitute $\bf W$ in \eqref{eq_33} into the power constraint, and we obtain
\begin{equation}
\begin{aligned}
& \left\| {{\bf{Ws}}} \right\|_2^2 = p_0 \\
\Rightarrow & {\kern 3pt}{{\bf{s}}^H}{{\bf{W}}^H}{\bf{Ws}}=p_0 \\
\Rightarrow & {\kern 3pt} {{\bf{s}}_{\bf E}^H}diag\left( {\bf \Omega} \right){{\bf{U}}^H}{\left( {{\bf{H}}{{\bf{H}}^H}} \right)^{ - 1}}{\bf{U}}diag\left( {\bf \Omega} \right){\bf{s}}_{\bf E} = p_0 \\
\Rightarrow & {\kern 3pt} {\bf \Omega}^T \underbrace {diag\left( {{{{\bf{s}}}_{\bf E}^H}} \right){{\bf{U}}^H}{{\left( {{\bf{H}}{{\bf{H}}^H}} \right)}^{ - 1}}{\bf{U}}diag\left( {{\bf{s}}}_{\bf E} \right)}_{{\bf{T}}} {\bf \Omega} = p_0 \\
\Rightarrow & {\kern 3pt} {\bf \Omega}^T {\bf T}  {\bf \Omega}=p_0.
\end{aligned}
\label{eq_34}
\end{equation}
Since ${\bf T}$ is Hermitian and positive semi-definite, and since each entry in $\bf \Omega$ is real, \eqref{eq_34} can be further transformed into
\begin{equation}
{\bf \Omega}^T {\bf T}  {\bf \Omega} = {\bf \Omega}^T \Re \left\{{\bf T} \right\} {\bf \Omega} = {\bf \Omega}^T {\bf V}  {\bf \Omega} = p_0,
\label{eq_35}
\end{equation}
where ${\bf V} = \Re \left\{{\bf T} \right\}$ is symmetric. With the expression for $\bf W$ in \eqref{eq_33} and the updated power constraint, we are able to construct an equivalent optimization on $\bf \Omega$, given by
\begin{equation}
\begin{aligned}
&\mathcal{P}_5: {\kern 3pt} \mathop {\min }\limits_{{{\bf \Omega}}, {\kern 1pt} t} {\kern 3pt} -t \\
&{\kern 0pt} s. t. {\kern 10pt} {\bf \Omega}^T {\bf V}  {\bf \Omega} - p_0 = 0 \\
&{\kern 24pt} t - \alpha_m^{\cal O} \le 0, {\kern 3pt} \forall \alpha_m^{\cal O} \in {\cal O}\\
&{\kern 24pt} t - \alpha_n^{\cal I} = 0 , {\kern 3pt} \forall \alpha_n^{\cal I} \in {\cal I}
\label{eq_36}
\end{aligned}
\end{equation}
The optimal precoding matrix for the original optimization ${\cal P}_3$ is then obtained by substituting the solution of ${\cal P}_5$ into \eqref{eq_33}. In the following, we analyze ${\cal P}_5$ and derive the closed-form optimal precoding matrix as a function of the dual variables of ${\cal P}_5$. 

The Lagrangian of ${\cal P}_5$ is formulated as
\begin{equation}
\begin{aligned}
&{\cal L} \left( {{\bf \Omega} ,t,{\delta _0},{\mu _m},{\nu _n}} \right) =  - t + {\delta _0}\left( {{{\bf \Omega}^T}{\bf{V}}{\bf \Omega}  - {p_0}} \right)\\
& + \sum\limits_{m = 1}^{card\left\{ {\cal O} \right\}} {{\mu _m}\left( {t - \alpha _m^{\cal O}} \right)}  + \sum\limits_{n = 1}^{card\left\{ {\cal I} \right\}} {{\nu _n}\left( {t - \alpha _n^{\cal I}} \right)},
\end{aligned}
\label{eq_37}
\end{equation}
where $\mu_m \ge 0$, $\forall m \in \left\{ {1,2,\cdots, card\left\{ {\cal O} \right\}} \right\}$. To simplify the subsequent KKT conditions, we propose to reorder the columns and rows of the matrices and vectors included in the Lagrangian expression in \eqref{eq_37}. Specifically, we reorder the expanded symbol vector ${\bf s_E}$ into
\begin{equation}
{\bf s_E} \Rightarrow {\bf \tilde s_E}={\left[ {{\bf{\tilde s}}_{\cal O}^T, {\bf{\tilde s}}_{\cal I}^T } \right]^T},
\label{eq_38}
\end{equation}
where ${\bf{\tilde s}}_{\cal O} \in {\cal R}^{card\left\{ {\cal O} \right\} \times 1}$ and ${\bf{\tilde s}}_{\cal I} \in {\cal R}^{card\left\{ {\cal I} \right\} \times 1}$ are given by
\begin{equation}
\begin{aligned}
&{\bf{\tilde s}}_{\cal O} = {\left[ {\tilde s_1,\tilde s_2, \cdots ,\tilde s_{card\left\{ {\cal O} \right\}}} \right]^T}, \\
&{\bf{\tilde s}}_{\cal I} = \left[{\tilde s_{card\left\{ {\cal O} \right\} + 1},\tilde s_{card\left\{ {\cal O} \right\} + 2}, \cdots ,\tilde s_{2K}} \right]^T,
\end{aligned}
\label{eq_39}
\end{equation}
such that the entries in ${\bf{\tilde s}}_{\cal O}$ correspond to the real or imaginary part of the outer constellation points that can exploit CI, and the entries in ${\bf{\tilde s}}_{\cal I}$ correspond to the real and imaginary part of the inner constellation points that cannot be scaled. The corresponding scaling vector $\bf \Omega$ is accordingly transformed into
\begin{equation}
{\bf \Omega} \Rightarrow {\bf \tilde \Omega}= {\left[ { {\bf \tilde \Omega}_{\cal O}^T, {\bf \tilde \Omega}_{\cal I}^T} \right]^T},
\label{eq_40}
\end{equation}
where ${\bf \tilde \Omega}_{\cal O} \in {\cal R}^{card\left\{ {\cal O} \right\} \times 1}$ and ${\bf \tilde \Omega}_{\cal I} \in {\cal R}^{card\left\{ {\cal I} \right\} \times 1}$ are given by
\begin{equation}
\begin{aligned}
&{\bf \tilde \Omega}_{\cal O} = {\left[ {\tilde \alpha_1,\tilde \alpha_2, \cdots ,\tilde \alpha_{card\left\{ {\cal O} \right\}}} \right]^T}, \\
&{\bf \tilde \Omega}_{\cal I} = \left[{\tilde \alpha_{card\left\{ {\cal O} \right\} + 1}, \tilde \alpha_{card\left\{ {\cal O} \right\} + 2}, \cdots ,\tilde \alpha_{2K}} \right]^T.
\end{aligned}
\label{eq_41}
\end{equation}

We further introduce a `{\bf Locater}' function that returns the index of ${\tilde s}_m$ in the original expanded symbol vector ${\bf s_E}$, given by
\begin{equation} 
{\rm{L}}\left( {{{\tilde s}_m}} \right) = k, {\kern 3pt} {\rm{if}} {\kern 3pt} {\tilde s_m} = s_k^{E}.
\label{eq_42}
\end{equation}
We can then express ${\bf \tilde s_E}$ and $\bf \tilde \Omega$ as
\begin{equation}
{\bf \tilde s_E} = {\bf F}{\bf s_E}, {\kern 3pt} {\bf \tilde \Omega} = {\bf F} {\bf \Omega},
\label{eq_43}
\end{equation}
where the transformation matrix ${\bf F} \in {\cal R}^{2K \times 2K}$ that transforms the original $\bf \Omega$ and ${\bf s_E}$ into their reordered forms is given by
\begin{equation}
{\bf F} = {\left[ {{\bf{e}}_{{\rm{L}}\left( {\tilde s_1} \right)},{\bf{e}}_{{\rm{L}}\left( {\tilde s_2} \right)}, \cdots ,{\bf{e}}_{{\rm{L}}\left( {\tilde s_{2K}} \right)}} \right]^T},
\label{eq_44}
\end{equation}
and we note that ${\bf F}$ is invertible. Similarly, the corresponding reordered matrix ${\bf V}$ can be obtained as
\begin{equation}
{\bf \tilde V} = {\bf F}{\bf V}{\bf F}^T,
\label{eq_45}
\end{equation} 
where the multiplication of ${\bf F}$ at the left side and ${{\bf{F}}^T}$ at the right side correspond to the row and column reordering, respectively. Using the above expressions for $\bf \tilde s$, $\bf \tilde \Omega$ and ${\bf \tilde V}$, the Lagrangian of ${\cal P}_5$ in \eqref{eq_37} can be further transformed into a simple form, given by
\begin{equation}
{\cal L} \left( {{\bf \tilde \Omega} ,t,{\delta _0}, {\bf u_1}} \right) = \left( {{{\bf{1}}^T}{\bf {u_1}} - 1} \right)t + {\delta _0} \cdot {\bf \tilde \Omega} ^T{\bf \tilde V} {\bf \tilde \Omega} - {{\bf {u}}_{\bf 1}^T}{\bf \tilde \Omega}  - {\delta _0}{p_0},
\label{eq_46}
\end{equation}
where ${\bf 1}=\left[ {1,1,\cdots,1}\right]^T \in {\cal R}^{2K \times 1}$, and ${\bf u_1} \in {\cal R}^{2K \times 1}$ is the dual vector corresponding to the reordered $\bf \tilde \Omega$, given by
\begin{equation}
{\bf {u_1}} = \left[ {{\mu _1},{\mu _2}, \cdots ,{\mu _{card\left\{ {\cal O} \right\}}},{\nu _1},{\nu _2}, \cdots ,{\nu _{card\left\{ {\cal I} \right\}}}} \right]^T.
\label{eq_47}
\end{equation}
Subsequently, the KKT conditions for ${\cal P}_5$ can be formulated as
\begin{IEEEeqnarray}{rCl} 
\IEEEyesnumber
\frac{{\partial {\cal L}}}{{\partial t}} ={\bf 1}^T {\bf u_1}  - 1    = 0 {\kern 30pt} \IEEEyessubnumber* \label{eq_48a} \\
\frac{{\partial \cal L}}{{\partial \bf \tilde \Omega}} = {\delta _0} \cdot 2 {\bf{\tilde V}} {\bf \tilde \Omega}  - {\bf {u_1}} = {\bf 0} {\kern 30pt} \label{eq_48b} \\
{\bf \tilde \Omega} ^T{\bf{\tilde V}}{\bf \tilde \Omega} - p_0 = 0 {\kern 30pt} \label{eq_48c} \\
{\mu _m}\left( {t - {\tilde \alpha_m}} \right) = 0, {\kern 2pt} \forall m \in \left\{ {1,2,\cdots, card\left\{ {{\cal O}} \right\}} \right\} {\kern 28pt} \label{eq_48d}\\
t - {\tilde \alpha _n} = 0, {\kern 2pt} \forall n \in \left\{ {card\left\{ {{\cal I}} \right\}+1,\cdots, 2K} \right\} {\kern 28pt} \label{eq_48e}
\end{IEEEeqnarray}
Based on \eqref{eq_48b}, we obtain an expression for $\bf \tilde \Omega$ as a function of $\bf u_1$, given by
\begin{equation}
{\bf \tilde \Omega} = \frac{1}{{2{\delta _0}}} \cdot {{\bf{\tilde V}}^{ - 1}}{\bf {u_1}},
\label{eq_49}
\end{equation}
where we note that ${\bf \tilde V}$ is symmetric and invertible. By substituting the expression for ${\bf \tilde \Omega}$ in \eqref{eq_49} into the power constraint, we further obtain $\delta_0$ as
\begin{equation}
\begin{aligned}
&{\left( {\frac{1}{{2{\delta _0}}} \cdot {{{\bf{\tilde V}}}^{ - 1}}{\bf {u_1}}} \right)^T}{{\bf{\tilde V}}}\left( {\frac{1}{{2{\delta _0}}} \cdot {{{\bf{\tilde V}}}^{ - 1}}{\bf {u_1}}} \right) = {p_0} \\
\Rightarrow & {\kern 2pt} \frac{1}{{4\delta _0^2}} \cdot {{\bf {u}}_{\bf 1}^T}{{\bf{\tilde V}}^{ - 1}}{\bf{\tilde V}}{{\bf{\tilde V}}^{ - 1}}{\bf {u_1}} = {p_0}\\
\Rightarrow & {\kern 2pt} {\delta _0} = \sqrt {\frac{{{{\bf {u}}_{\bf 1}^T}{{{\bf{\tilde V}}}^{ - 1}}{\bf {u_1}}}}{{4{p_0}}}}.
\end{aligned}
\label{eq_50}
\end{equation}

For the convex optimization ${\cal P}_5$, it is easy to verify that Slater's condition is met \cite{r27}, which means that the dual gap is zero. Accordingly, ${\cal P}_5$ can also be optimally solved via its dual problem, given by
\begin{equation}
\mathcal{P}_6: {\kern 3pt} {\cal D}= \mathop {\max }\limits_{{\bf {u_1}}, {\delta _0}} \mathop {\min }\limits_{{\bf \tilde \Omega} , t} {\cal L}\left( {{\bf {u_1}},{\delta _0},{\bf \tilde \Omega} ,t} \right).
\label{eq_51}
\end{equation}
For the above dual problem, the inner minimization is achieved by \eqref{eq_48a} and \eqref{eq_49}, and we can further simplify the dual problem into
\begin{equation}
\begin{aligned}
{\cal D}&=\mathop {\max }\limits_{{\bf {u_1}},{\delta _0}} {\kern 2pt} {\delta _0} \cdot {\bf \tilde \Omega} ^T{\bf{\tilde V}}{\bf \tilde \Omega} + {{\bf {u}}_{\bf 1}^T}{\bf \tilde \Omega}  - {\delta _0}{p_0}\\
&=\mathop {\max }\limits_{{\bf {u_1}},{\delta _0}} {\kern 2pt} \frac{{{\delta _0}}}{{4\delta _0^2}} \cdot {{\bf {u}}_{\bf 1}^T}{{\bf{\tilde V}}^{ - 1}}{\bf {u_1}}- \frac{1}{{2{\delta _0}}} {{\bf {u}}_{\bf 1}^T}{{\bf{\tilde V}}^{ - 1}}{\bf  {u_1}} - {\delta _0}{p_0}\\
&=\mathop {\max }\limits_{{\bf {u_1}}} {\kern 2pt}  - \frac{{{{\bf {u}}_{\bf 1}^T}{{{\bf{\tilde V}}}^{ - 1}}{\bf {u_1}}}}{{4\sqrt {\frac{{{{\bf {u}}_{\bf 1}^T}{{{\bf{\tilde V}}}^{ - 1}}{\bf {u_1}}}}{{4{p_0}}}} }} - \sqrt {\frac{{{{\bf {u}}_{\bf 1}^T}{{{\bf{\tilde V}}}^{ - 1}}{\bf {u_1}}}}{{4{p_0}}}}  \cdot {p_0}\\
&=\mathop {\max }\limits_{{\bf {u_1}}} {\kern 2pt}  - \sqrt {{p_0} \cdot {{\bf {u}}_{\bf 1}^T}{{{\bf{\tilde V}}}^{ - 1}}{\bf {u_1}}} 
\end{aligned}
\label{eq_52}
\end{equation}
Based on the fact that $y=\sqrt x$ is a monotonic function, the above dual problem is equivalent to the following minimization problem:
\begin{equation}
\begin{aligned}
&\mathcal{P}_7: {\kern 3pt} \mathop {\min }\limits_{{{\bf u_1}}} {\kern 3pt} {\bf u}_{\bf 1}^T {\bf \tilde V}^{-1} {\bf u_1} \\
&{\kern 0pt} s. t. {\kern 10pt} {\bf 1}^T{\bf u_1} - 1 = 0 \\
&{\kern 24pt} \mu_m \ge 0, {\kern 3pt} \forall m \in \left\{ {1,2,\cdots,card\left\{ {\cal O} \right\}} \right\}
\label{eq_53}
\end{aligned}
\end{equation}
which is a QP optimization and can be more efficiently solved than the SOCP formulation. Moreover, based on the expression for $\bf \tilde \Omega$ in \eqref{eq_49} and $\delta_0$ in \eqref{eq_50}, we finally obtain the optimal closed-form precoding matrix $\bf W$ as a function of the dual vector $\bf u_1$ in the case of $K \le N_t$ as
\begin{equation}
\begin{aligned}
&{\bf{W}} =\\
& \frac{1}{K} {{\bf{H}}^H}{\left( {{\bf{H}}{{\bf{H}}^H}} \right)^{ - 1}}{\bf{U}}diag\left( {\sqrt {\frac{{{p_0}}}{{{{\bf {u}}_{\bf 1}^T}{{{\bf{\tilde V}}}^{ - 1}}{\bf {u_1}}}}} {\bf{F}}^{-1} {{{\bf{\tilde V}}}^{ - 1}}{\bf {u_1}}} \right){{\bf{s}}_{\bf E}}{{\bf{\hat s}}^T},
 \end{aligned}
\label{eq_54}
\end{equation}
where ${\bf F}^{-1}$ is to order the obtained $\bf \tilde \Omega$ into the original $\bf \Omega$, with ${\bf F}$ given in \eqref{eq_44}. 

Compared to the final QP formulation for PSK modulation in \cite{r45} that is optimized over a simplex, a key difference for the case of QAM constellations is that the variable vector is no longer on a simplex, and only the dual variables that correspond to the real and imaginary part of the constellation points that can exploit CI are constrained to be non-negative, as observed in ${\cal P}_7$. We note that both QP formulations for PSK and QAM modulations can be solved by convex optimization tools. However, for the reasons given above, the more efficient simplex method that is generally used for solving QP problems over a simplex and the proposed iterative algorithm in \cite{r45} are not directly applicable to such multi-level modulations.

\section{CI Precoding for the Case of $K > N_t$}
In this section, we further extend our study to the case where the BS simultaneously serves a number of users larger than the number of the transmit antennas at the BS, i.e., $K>N_t$. Specifically, our derivations in this section and the corresponding numerical results show that, by exploiting the information of the channel as well as the data symbols and by judiciously constructing the precoding matrix, CI precoding is able to spatially multiplex more data streams than the number of transmit antennas. Similar to the case of $K \le N_t$, the subsequent analysis is generic and can be further extended to other multi-level constellations.

When $K > N_t$, the direct inverse included in \eqref{eq_32} becomes infeasible, as the product ${\bf HH}^H$ is rank-deficient. In this case, the more general pseudo inverse instead of the direct matrix inverse is employed \cite{r46}. Based on \eqref{eq_31}, we can now express ${\bf{\Upsilon}}$ in the case of $K > N_t$ as
\begin{equation}
{\bf{\Upsilon}}=\frac{1}{K} \cdot {\left( {{\bf{H}}{{\bf{H}}^H}} \right)^ + }{\bf{U}}diag\left( \bf \Omega  \right){\bf s_E},
\label{eq_55}
\end{equation}
and the obtained precoding matrix $\bf W$ as
\begin{equation}
{\bf W}=\frac{1}{K} \cdot {{\bf{H}}^H}{\left( {{\bf{H}}{{\bf{H}}^H}} \right)^{+}}{\bf{U}}diag\left( {\bf \Omega} \right){\bf s_E}{{\bf{\hat s}}^T}.
\label{eq_56}
\end{equation}
By substituting the expression for the obtained precoding matrix $\bf W$ into the power constraint, we can similarly obtain
\begin{equation}
{{\bf \Omega} ^T}diag\left( {{\bf{s}}_{\bf{E}}^H} \right){{\bf{U}}^H}{\left( {{\bf{H}}{{\bf{H}}^H}} \right)^ + }{\bf{U}}diag\left( {{{\bf{s}}_{\bf{E}}}} \right){\bf \Omega}  = {p_0}.
\label{eq_57}
\end{equation}
Then, one can easily follow a similar approach to that in Section III to obtain a QP optimization and the corresponding solution. However, we note that the solution obtained by following the above procedure is not a valid one for the original problem, since the inclusion of the pseudo inverse does not guarantee the equality of the original constraint. To be more specific, if we consider $K \le N_t$ and substitute the obtained precoding matrix in \eqref{eq_33} into \eqref{eq_28}, we obtain
\begin{equation}
\begin{aligned}
&{\bf{HWs}} = {\bf{U}}diag\left( \bf \Omega  \right){\bf s_E} \\
\Rightarrow & {\bf{H}} \left[{ \frac{1}{K} {{\bf{H}}^H}{\left( {{\bf{H}}{{\bf{H}}^H}} \right)^{ - 1}}{\bf{U}}diag\left( \bf \Omega  \right){\bf s_E}{{\bf{\hat s}}^T}}\right]{\bf{s}} = {\bf{U}}diag\left( \bf \Omega  \right){\bf s_E}\\
\Rightarrow & {\bf{U}}diag\left( \bf \Omega  \right){\bf s_E} = {\bf{U}}diag\left( \bf \Omega  \right){\bf s_E},
\end{aligned}
\label{eq_58}
\end{equation}
which is always true. This means that the symbol-scaling constraint in \eqref{eq_28} is already implicitly included in the power constraint in \eqref{eq_34} for the case of $K \le N_t$. However, in the case of $K > N_t$ where the pseudo inverse is employed, the above equality may not hold and simply following the approach for $K \le N_t$ will lead to an erroneous solution. Therefore, the following additional constraint is further required in the case of $K > N_t$ to obtain a valid and correct solution:
\begin{equation}
\begin{aligned}
&{\kern 3pt} {\bf{H}} \left[{\frac{1}{K} {{\bf{H}}^H}{\left( {{\bf{H}}{{\bf{H}}^H}} \right)^{+}}{\bf{U}}diag\left( \bf \Omega  \right){\bf s_E}{{\bf{\hat s}}^T}}\right]{\bf{s}} = {\bf{U}}diag\left( \bf \Omega  \right){\bf s_E}\\
\Rightarrow & {\kern 3pt} {\bf{H}}{{\bf{H}}^H}{\left( {{\bf{H}}{{\bf{H}}^H}} \right)^ + }{\bf{U}}diag\left( \bf \Omega  \right){\bf s_E} = {\bf{U}}diag\left( \bf \Omega  \right){\bf s_E}\\
\Rightarrow & \left[ {{\bf{H}}{{\bf{H}}^H}{{\left( {{\bf{H}}{{\bf{H}}^H}} \right)}^ + } - {{\bf{I}}_K}} \right]{\bf{U}}diag\left( \bf \Omega  \right){\bf s_E} = {\bf{0}}\\
\Rightarrow & \underbrace {\left[ {{\bf{H}}{{\bf{H}}^H}{{\left( {{\bf{H}}{{\bf{H}}^H}} \right)}^ + } - {{\bf{I}}_K}} \right]{\bf{U}}diag\left( {\bf s_E}\right)}_{\bf{P}}{\bf \Omega} = {\bf{0}},
\end{aligned}
\label{eq_59}
\end{equation}
where the matrix $\mathbf{P}$ satifies the following property.

{\bf Observation 2}: The rank of the coefficient matrix $\bf P$ is $\left( {K-N_t} \right)$ with probability 1.

{\bf Proof}: We first consider the matrix ${\bf A}={\bf{H}}{{\bf{H}}^H}{{\left( {{\bf{H}}{{\bf{H}}^H}} \right)}^ + }$, and we note that ${\rm{rank}}\left\{ {\bf A} \right\}=N_t$ with probability 1. Accordingly, we can express the eigenvalue decomposition of $\bf A$ as
\begin{equation}
{\bf{A}} = {\bf{Q}}{\bf \Lambda _A}{{\bf{Q}}^H},
\end{equation}
where ${\bf{Q}}$ is a unitary matrix. Based on the construction of the pseudo-inverse \cite{r46}, $\bf \Lambda _A$ is given by
\begin{equation}
{\bf \Lambda _A}=diag\left\{ {{{\left[ {1, \cdots ,1,0, \cdots ,0} \right]}^T}} \right\},
\end{equation}
with all the $N_t$ non-zero eigenvalues equal to 1. By further expressing the identity matrix ${\bf I}_K$ as
\begin{equation}
{{\bf{I}}_K} = {\bf{Q}}{\bf I}_K{{\bf{Q}}^H},
\end{equation}
we define
\begin{equation}
\begin{aligned}
{\bf B}&={\bf{H}}{{\bf{H}}^H}{\left( {{\bf{H}}{{\bf{H}}^H}} \right)^ + } -{{\bf{I}}_K}\\
&= {\bf{A}} - {{\bf{I}}_K}\\
&={\bf{Q}}\left( {{\bf \Lambda _A} - {{\bf{I}}_K}} \right){{\bf{Q}}^H}\\
&={\bf{Q}}{\bf \Lambda _B}{{\bf{Q}}^H},
\end{aligned}
\end{equation}
where the last step represents the eigenvalue decomposition of $\bf B$. Since ${\bf \Lambda _B}$ contains $\left( {K-N_t} \right)$ non-zero entries on the diagonal, we obtain that ${\rm{rank}}\left\{ {\bf{B}} \right\} = K - {N_t}$. Subsequently, based on the observation that both $\bf U$ and $diag \left( {\bf s_E} \right)$ are full-rank, we therefore have
\begin{equation}
{\rm{rank}}\left\{ {\bf P} \right\}={\rm{rank}}\left\{ {\bf B} \right\}=K-N_t
\end{equation}
with probability 1, and the proof is complete. ${\kern 55pt} \blacksquare$

\newcounter{mytempeqncnt}
\begin{figure*}
\normalsize
\setcounter{mytempeqncnt}{\value{equation}}
\setcounter{equation}{74}

\begin{equation}
{\bf W}= \frac{1}{K} \cdot {{\bf{H}}^H}{\left( {{\bf{H}}{{\bf{H}}^H}} \right)^{ +}}{\bf U} diag\left( { \sqrt {\frac{{{p_0}}}{{{\bf{u}}_{\bf{2}}^T{{\bf{F}}}{{\bf{D}}}{\bf{Y}}^{ - 1}{\bf{D}}^T{\bf{F}}^T{{\bf{u}}_{\bf{2}}}}}}  \cdot {{\bf{D}}}{\bf{Y}}^{ - 1}{\bf{D}}^T{\bf{F}}^T{{\bf{u}}_{\bf{2}}} } \right){\bf{s_E}}{{\bf{\hat s}}^T},
\label{eq_70}
\end{equation}

\setcounter{equation}{\value{mytempeqncnt}}
\hrulefill
\vspace*{4pt}
\end{figure*}

Based on \eqref{eq_59}, while it is obvious that ${\bf \Omega}={\bf 0}$ can be the solution of the above equation set, this is not a valid solution to the original CI problem. Therefore, this additional constraint is equivalent to having non-zero solutions $\bf \Omega$ for the linear equation ${\bf P \Omega} = {\bf 0}$. Noting that $\bf P$ is complex while $\bf \Omega$ is constrained to be real, we expand ${\bf P} \in {\cal C}^{K \times 2K}$ into its real equivalent ${\bf P_E} \in {\cal R}^{2K \times 2K}$, given by
\begin{equation}
{\bf P_E} = \left[ {\begin{array}{*{20}{c}}
{\Re \left( {\bf{P}} \right)}\\
{\Im \left( {\bf{P}} \right)}
\end{array}} \right],
\label{eq_60}
\end{equation}
and based on {\bf Observation 2} we obtain that ${\rm{rank}}\left\{ {{{\bf{P}}_{\bf{E}}}} \right\}=2\left({K-N_t} \right)$ with probability 1. We further express the singular value decomposition (SVD) of $\bf P_E$ as
\begin{equation}
{{\bf{P}}_{\bf{E}}} = {{\bf{S}}}{{\bf \Sigma}}{\bf \hat{D}}^H,
\label{eq_61}
\end{equation}
where ${\bf \hat D} = \left[ {{{\bf \hat{d}}_1},{{\bf \hat{d}}_2}, \cdots ,{{\bf \hat{d}}_{2K}}} \right]$ is the matrix that consists of the right singular vectors. Thus, the non-zero solution $\bf \Omega$ is in the null space of ${\bf P_E}$, and can be expressed as a linear combination of the right singular vectors that correspond to zero singular values:
\begin{equation}
\begin{aligned}
{\bf \Omega}  &= \sum\limits_{n = 1}^{2K - {\rm{rank}}\left\{ {{{\bf{P}}_{\bf{E}}}} \right\}} {{\beta_n} \cdot {{\bf \hat{d}}_{{\rm{rank}}\left\{ {{{\bf{P}}_{\bf{E}}}} \right\} + n}}}  \\
&=\sum\limits_{n = 1}^{2{N_t}} {{\beta _n} \cdot {{{\bf{\hat d}}}_{2\left( {K - {N_t}} \right) + n}}} \\
&= {{\bf{D}}}{\bm{\beta}},
\end{aligned}
\label{eq_62}
\end{equation}
where each $\beta_n$ is real and represents the weight for the corresponding singular vector, ${\bm \beta} = \left[ {\beta_1,\beta_2,\cdots,\beta_{2N_t}} \right]^T$, and ${\bf D} \in {\cal R}^{2K \times 2N_t}$ consists of the right singular vectors of $\bf P_E$ that correspond to the zero singular values, given by
\begin{equation}
\begin{aligned}
{\bf D} &= \left[ {{{\bf \hat{d}}_{ 2\left( {K-N_t} \right) + 1}},{{\bf \hat{d}}_{ 2\left( {K-N_t} \right)+ 2}}, \cdots ,{{\bf \hat{d}}_{2K}}} \right] \\
&= \left[ {{{\bf{d}}_1},{{\bf{d}}_2}, \cdots ,{{\bf{d}}_{2K}}} \right]^T,
\end{aligned}
\label{eq_63}
\end{equation}
where each ${\bf d}_k^T$ then represents the $k$-th row of $\bf D$. By substituting the expression for $\bf D$ into the power constraint in \eqref{eq_57}, we further obtain
\begin{equation}
{{\bm{\beta}}^T}\underbrace {{\bf{D}}^Tdiag\left( {{\bf{s}}_{\bf{E}}^{{H}}} \right){{\bf{U}}^H}{{\left( {{\bf{H}}{{\bf{H}}^H}} \right)}^ + }{\bf{U}}diag\left( {{{\bf{s}}_{\bf{E}}}} \right){{\bf{D}}}}_{{{\bf{X}}}}{\bm{\beta}} = {p_0},
\label{eq_64}
\end{equation}
which is the valid power constraint for the case of $K>N_t$. The reason for the key difference in the power constraint between the case of $K \le N_t$ and $K > N_t$ is that, while the symbol-scaling constraint in \eqref{eq_28} is automatically satisfied by \eqref{eq_34} for the case of $K \le N_t$, it may not hold for the case of $K > N_t$ and therefore the expression for $\bf \Omega$ in \eqref{eq_62} is required to guarantee the symbol-scaling constraint \eqref{eq_28} is met for the original CI precoding.

Based on the above analysis, we have obtained an expression for $\bf \Omega$ as a function of $\bm \beta$. Subsequently, we can now formulate an equivalent optimization on $\bm \beta$, given by
\begin{equation}
\begin{aligned}
&\mathcal{P}_{11}: {\kern 3pt} \mathop {\min }\limits_{{{\bm \beta}}, {\kern 1pt} t} {\kern 3pt} -t \\
&{\kern 2pt} s. t. {\kern 13pt} {\bm \beta}^T {\bf Y}  {\bm \beta} - p_0 = 0 \\
&{\kern 29pt} t - {\bf{d}}_m^T{\bm{\beta}} \le 0, {\kern 3pt} \forall \alpha_m^{\cal O} \in {\cal O}\\
&{\kern 29pt} t - {\bf{d}}_n^T{\bm{\beta}} = 0 , {\kern 3pt} \forall \alpha_n^{\cal I} \in {\cal I}
\label{eq_65}
\end{aligned}
\end{equation}
where ${{\bf{Y}}} = \Re \left( {{{\bf{X}}}} \right)$. The Lagrangian of ${\cal P}_{11}$ can be expressed as
\begin{equation}
\begin{aligned}
{\cal L}= &- t + {\tilde \delta _0}\left( {{{\bm \beta} ^T}{\bf{Y}}{\bm \beta}  - {p_0}} \right) + \sum\limits_{m = 1}^{card\left\{ \cal O \right\}} {{{\tilde \mu }_m}\left( {t - {{\bf{d}}_m^T}{\bm \beta} } \right)} \\
& + \sum\limits_{n = 1}^{card\left\{ \cal I \right\}} {{{\tilde \nu }_n}\left( {t - {{\bf{d}}_n^T}{\bm \beta} } \right)} \\
=&\left( {{{\bf{1}}^T}{{\bf{u}}_{\bf{2}}} - 1} \right)t + {\tilde \delta _0} \cdot {{\bm \beta} ^T}{\bf{Y}}{\bm \beta}  - {\bf{u}}_{\bf{2}}^T{\bf{FD}}{\bm \beta}  - {\tilde \delta _0}{p_0},
\end{aligned}
\label{eq_66}
\end{equation}
where the dual vector $\bf u_2$ for ${\cal P}_{11}$ is given by
\begin{equation}
{\bf {u_2}} = \left[ {{\tilde \mu _1},{\tilde \mu _2}, \cdots ,{\tilde \mu _{card\left\{ {\cal O} \right\}}},{\tilde \nu _1},{\tilde \nu _2}, \cdots ,{\tilde \nu _{card\left\{ {\cal I} \right\}}}} \right]^T.
\label{eq_67}
\end{equation}
The corresponding KKT conditions are:
\begin{IEEEeqnarray}{rCl} 
\IEEEyesnumber
\frac{{\partial {\cal L}}}{{\partial t}} = {{\bf{1}}^T}{{\bf{u}}_{\bf{2}}} - 1 = 0 {\kern 30pt} \IEEEyessubnumber* \label{eq_68a} \\
\frac{{\partial {\cal L}}}{{\partial \bm \beta }} = {\tilde \delta _0} \cdot 2{\bf Y} {\bm \beta}  - {{\bf{D}}^T}{{\bf{F}}^T}{{\bf{u}}_{\bf{2}}} = {\bf{0}} {\kern 30pt} \label{eq_68b} \\
{\bm \beta}^T {\bf Y}  {\bm \beta} - p_0 = 0 {\kern 30pt} \label{eq_68c} \\
\tilde \mu_m \left({t - {\bf{d}}_m^T{\bm{\beta}}}\right) = 0, {\kern 3pt} \forall \alpha_m^{\cal O} \in {\cal O} {\kern 30pt} \label{eq_68d} \\
t - {\bf{d}}_n^T{\bm{\beta}} = 0 , {\kern 3pt} \forall \alpha_n^{\cal I} \in {\cal I} {\kern 30pt} \label{eq_68e}
\end{IEEEeqnarray}
Then, by following a similar approach to that in Section III, the dual problem of ${\cal P}_{11}$ can be finally formulated into a QP optimization as well, given by
\begin{equation}
\begin{aligned}
&\mathcal{P}_{12}: {\kern 3pt} \mathop {\min }\limits_{{{\bf u_2}}} {\kern 3pt} {\bf{u}}_{\bf{2}}^T\left( { {{\bf{F}}}{{\bf{D}}}{\bf{Y}}^{ - 1}{\bf{D}}^T{\bf{F}}^T } \right){{\bf{u}}_{\bf{2}}} \\
&{\kern 2pt} s. t. {\kern 13pt} {\bf 1}^T{\bf u_2} - 1 = 0 \\
&{\kern 29pt} \tilde \mu_m \ge 0, {\kern 3pt} \forall m \in \left\{ {1,2,\cdots,card\left\{ {\cal O} \right\}} \right\}
\label{eq_69}
\end{aligned}
\end{equation}
where we have omitted the details for brevity. The optimal precoding matrix can then be expressed in closed form as a function of the dual vector ${\bf u}_2$, which is shown in \eqref{eq_70} at the top of this page. 

\subsection{The Condition for Multiplexing $K>N_t$ Streams}
Based on ${\cal P}_{11}$ and the above analysis, we can also obtain an expression for the optimal $t$ in ${\cal P}_{11}$ as
\setcounter{equation}{75}
\begin{equation}
t^* = \mathop {\min }\limits_k {\kern 2pt} {\bf{d}}_k^T{\bm{\beta}}, {\kern 3pt} \forall k \in \left\{ {1,2,\cdots,2K} \right\}.
\label{eq_71}
\end{equation}
Accordingly, we obtain that if $t^*>0$, the obtained scaling vector $\bf \Omega$ and the resulting precoding matrix is a valid solution to the original CI precoding. Otherwise if $t^*<0$, each data symbol will be scaled to the other 3 quarters of the constellation, which will lead to erroneous demodulation. Therefore, the condition
\begin{equation}
\mathop {\min }\limits_k \left( {{{\bf{d}}_k^T} {\bm \beta} } \right) > 0
\label{eq_72}
\end{equation} 
determines when multiplexing $K>N_t$ users is achievable.

\section{A Generic Iterative Algorithm for Multi-Level Modulations}
Since not all the variables are constrained to be non-negative in the QP formulation for multi-level modulations, as observed in ${\cal P}_7$ and ${\cal P}_{12}$, the iterative algorithm designed for PSK modulation does not directly apply to the case of multi-level constellations. Therefore in this section, we propose an iterative algorithm for a generic QP optimization, as an extension of the algorithm designed only for PSK modulations. Specifically, we focus on the following generic QP optimization
\begin{equation}
\begin{aligned}
&\mathcal{P}_{13}: {\kern 3pt} \mathop {\min }\limits_{{{\bf u}}} {\kern 3pt} {\bf{u}}^T {\bf Q} {{\bf{u}}} \\
&{\kern 2pt} s. t. {\kern 13pt} {\bf 1}^T{\bf u} - 1 = 0 \\
&{\kern 29pt} \mu_n \ge 0, {\kern 3pt} \forall n \in \left\{ {1,2,\cdots,N} \right\}
\label{eq_73}
\end{aligned}
\end{equation}
where $\bf Q$ is symmetric and positive definite, ${\bf u}=\left[ {\mu_1,\mu_2,\cdots,\mu_M} \right]^T \in {\cal C}^{M \times 1}$ and $N < M$. The above QP formulation can be regarded as a generalization of the QP formulation for both the case of $K\le N_t$ and $K>N_t$ considered in this paper, as well as for PSK modulations if $M=N$. We express the Lagrangian of ${\cal P}_{13}$ as
\begin{equation}
\begin{aligned}
{\cal L}&={{\bf{u}}^T}{\bf{Qu}} + {q_0}\left( {{{\bf{1}}^T}{\bf{u}} - 1} \right) - \sum\limits_{n = 1}^N {{q_n}{\mu _n}} \\
&={{\bf{u}}^T}{\bf{Qu}} + q_0 \cdot {{{\bf{1}}^T}{\bf{u}}}-{\bf{q}}_{\bf{E}}^T{\bf{u}}-q_0,
\end{aligned}
\label{eq_74}
\end{equation}
where ${\bf q}_{\bf E} \in {\cal C}^{M \times 1}$ is given by
\begin{equation}
{\bf q}_{\bf E}={\left[ {{q_1},{q_2}, \cdots ,{q_N},0, \cdots ,0} \right]^T} = {\left[ {{{\bf{q}}^T},{{\bf{0}}^{1 \times \left( {M - N} \right)}}} \right]^T}.
\label{eq_75}
\end{equation}
The corresponding KKT conditions can be obtained as
\begin{IEEEeqnarray}{rCl} 
\IEEEyesnumber
\frac{{\partial {\cal L}}}{{\partial {\bf{u}}}} = 2 {\bf Q} {\bf{u}} + {q_0} \cdot {\bf{1}} - {{\bf{q}}_{\bf{E}}} = 0 {\kern 30pt} \IEEEyessubnumber* \label{eq_76a} \\
{\bf 1}^T {\bf u}  - 1    = 0 {\kern 30pt} \label{eq_76b} \\
{q_n}{\mu _n} = 0, {\kern 2pt} \forall n \in \left\{ {1,2,\cdots,N} \right\} {\kern 27pt} \label{eq_76c} 
\end{IEEEeqnarray}
Based on the above, we obtain the expression for $\bf u$ as
\begin{equation}
{\bf u}=\frac{1}{2}{{\bf{Q}}^{ - 1}}\left( {{{\bf{q}}_{\bf{E}}} - {q_0} \cdot {\bf{1}}} \right)=\frac{1}{2}\left( {{{\bf{Q}}^{ - 1}}{{\bf{q}}_{\bf{E}}} - {q_0} \cdot {\bf{a}}} \right),
\label{eq_77}
\end{equation}
where ${\bf a}={\bf Q}^{-1} {\bf 1}$ represents the sum of all the columns of ${\bf Q}^{-1}$. By substituting the expression for $\bf u$ into \eqref{eq_76b}, we obtain $q_0$ as a function of ${\bf q_E}$:
\begin{equation}
\begin{aligned}
&{{\bf{1}}^T}\left[ {\frac{1}{2}\left( {{{\bf{Q}}^{ - 1}}{{\bf{q}}_{\bf{E}}} - {q_0} \cdot {\bf{a}}} \right)} \right] - 1 = 0\\
\Rightarrow &\frac{1}{2} \cdot {{\bf{1}}^T}{{\bf{Q}}^{ - 1}}{{\bf{q}}_{\bf{E}}} - \frac{{{q_0}}}{2} \cdot {{\bf{1}}^T}{\bf{a}} - 1 = 0\\
\Rightarrow & {q_0} = \frac{{{{\bf{1}}^T}{{\bf{Q}}^{ - 1}}{{\bf{q}}_{\bf{E}}} - 2}}{{{{\bf{1}}^T}{\bf{a}}}} \\
\Rightarrow & {q_0} = \frac{{{\bf{b}}{{\bf{q}}_{\bf{E}}} - 2}}{c},
\end{aligned}
\label{eq_78}
\end{equation}
where $c={\bf 1}^T{\bf a}$ denotes the sum of all the entries in ${\bf Q}^{-1}$, and ${\bf b}^T={\bf 1}^T{\bf Q}^{-1} $ is the sum of all the rows of ${\bf Q}^{-1}$. Due to the symmetry of $\bf Q$, we further obtain ${\bf b}={\bf a}^T$. By substituting $q_0$ into \eqref{eq_77}, we further obtain $\bf u$ as a function of $\bf q_E$:
\begin{equation}
\begin{aligned}
{\bf{u}} &= \frac{1}{2} \cdot {{\bf{Q}}^{ - 1}}{{\bf{q}}_{\bf{E}}} - \frac{{\bf{a}}}{2} \cdot \frac{{{\bf{b}}{{\bf{q}}_{\bf{E}}} - 2}}{c}\\
&=\frac{1}{2} \cdot {{\bf{Q}}^{ - 1}}{{\bf{q}}_{\bf{E}}} - \frac{{{\bf{ab}}{{\bf{q}}_{\bf{E}}}}}{{2c}} + \frac{{\bf{a}}}{c} \\
&=\frac{1}{2}\left( {{{\bf{Q}}^{ - 1}} - {\bf \Phi} } \right){{\bf{q}}_{\bf{E}}} + \frac{{\bf{a}}}{c},
\end{aligned}
\label{eq_79}
\end{equation}
where ${\bf \Phi}=\frac{{{\bf{ab}}}}{c}$. Based on the derived expression for $\bf u$, the following two observations can be made.

{\bf Observation 3}: When $K \le N_t$, ZF precoding can be viewed as a special case of CI-based precoding, when all the dual variables are forced to be zero.

{\bf Proof}: In this case, by comparing ${\cal P}_7$ with ${\cal P}_{13}$, we obtain that ${\bf Q}={\bf \tilde V}^{-1}$ and ${\bf a}={\bf \tilde V}{\bf 1}$. Then, based on the expression in \eqref{eq_49}, we can obtain $\bf \tilde \Omega$ as a function of $\bf q_E$:
\begin{equation}
\begin{aligned}
{\bf \tilde \Omega} &= \frac{1}{{2{\delta _0}}}{{\bf{\tilde V}}^{ - 1}}\left[ {\frac{1}{2}\left( {{\bf{\tilde V}} - {\bf \Phi} } \right){{\bf{q}}_{\bf{E}}} + \frac{{\bf{a}}}{c}} \right]\\
&=\frac{1}{{2{\delta _0}c}} \cdot {{{{\bf{\tilde V}}}^{ - 1}}{\bf{a}}} + \frac{1}{{4{\delta _0}}}\left( {{\bf{I}} - {{{\bf{\tilde V}}}^{ - 1}}{\bf \Phi}} \right){{\bf{q}}_{\bf{E}}}\\
&=\frac{1}{{2{\delta _0}c}} \cdot {{{{\bf{\tilde V}}}^{ - 1}}{{\bf{\tilde V1}}}} + \frac{1}{{4{\delta _0}}}\left( {{\bf{I}} - {{{\bf{\tilde V}}}^{ - 1}}{\bf \Phi}} \right){{\bf{q}}_{\bf{E}}}\\
&=\frac{1}{{2{\delta _0}c}} \cdot {\bf{1}} + \frac{1}{{4{\delta _0}}}\left( {{\bf{I}} - {{{\bf{\tilde V}}}^{ - 1}}\Phi } \right){{\bf{q}}_{\bf{E}}}.
\end{aligned}
\label{eq_80}
\end{equation}
When all the dual variables are zero, ${\bf q_E}={\bf 0}$, and we see that the scaling value in $\bf \tilde \Omega$ for each data symbol is identical, which then leads to the ZF precoder. ${\kern 120pt} \blacksquare$

{\bf Observation 4}: The solution for $\bf u$ in \eqref{eq_79} automatically satisfies ${\bf 1}^T{\bf u}-1=0$, irrespective of the value of ${\bf q_E}$.

{\bf Proof}: We calculate ${\bf 1}^T{\bf u}$, which can be expressed as
\begin{equation}
\begin{aligned}
{\bf 1}^T{\bf u} &= \frac{1}{2} \cdot {\bf 1}^T {{\bf{Q}}^{ - 1}}{{\bf{q}}_{\bf{E}}} - \frac{{{{\bf 1}^T\bf{ab}}{{\bf{q}}_{\bf{E}}}}}{{2c}} + \frac{{\bf 1}^T{\bf{a}}}{c} \\
&=\frac{1}{2} \cdot {\bf{b}}{{\bf{q}}_{\bf{E}}} - \frac{{c \cdot {\bf{b}}{{\bf{q}}_{\bf{E}}}}}{{2c}} + \frac{c}{c} \\
&=1,
\end{aligned}
\label{eq_81}
\end{equation}
which completes the proof. ${\kern 130pt} \blacksquare$

Following {\bf Observation 4}, ${\cal P}_{13}$ is equivalent to the following optimization problem:
\begin{equation}
\begin{aligned}
&\mathcal{P}_{14}: {\kern 3pt} {\rm{find}} {\kern 3pt} {\bf q_E} \\
&{\kern 2pt} s. t. {\kern 11pt} {\bf u}=\frac{1}{2}\left( {{{\bf{Q}}^{ - 1}} - {\bf \Phi} } \right){{\bf{q}}_{\bf{E}}} + \frac{{\bf{a}}}{c} \\
&{\kern 27pt} \mu_n q_n = 0, {\kern 3pt} \mu_n \ge 0, {\kern 3pt} q_n \ge 0, {\kern 3pt} \forall n \in \left\{ {1,2,\cdots,N} \right\}
\label{eq_82}
\end{aligned}
\end{equation}
Subsequently, based on the expression for $\bf q_E$ in \eqref{eq_75}, we decompose ${\bf Q}^{-1}$ and $\bf \Phi$ into 2 blocks, given by
\begin{equation}
{{\bf{Q}}^{ - 1}} = \left[ {\begin{array}{*{20}{c}}
{{{\bf{Q}}_{\bf{1}}}}&{{{\bf{Q}}_{\bf{2}}}}
\end{array}} \right], {\kern 3pt} {\bf \Phi}  = \left[ {\begin{array}{*{20}{c}}
{{{\bf \Phi} _{\bf{1}}}}&{{{\bf \Phi} _{\bf{2}}}}
\end{array}} \right],
\label{eq_83}
\end{equation}
where the dimension of both ${\bf Q_1}$ and $\bf \Phi_1$ is $M \times N$. Subsequently, $\bf u$ in \eqref{eq_79} can be further simplified and expressed as a function of $\bf q$, given by
\begin{equation}
\begin{aligned}
{\bf u}&=\frac{1}{2}\left( {{{\bf{Q_1}}} - {\bf \Phi_1} } \right){{\bf{q}}} + \frac{{\bf{a}}}{c}\\
&=\frac{1}{2} {\bf G}{{\bf{q}}} + \frac{{\bf{a}}}{c}.
\end{aligned}
\label{eq_84}
\end{equation}
Moreover, noting that only part of the entries in $\bf u$ are constrained to be non-negative in the original optimization ${\cal P}_{13}$, we further decompose $\bf u$, $\bf Q_1$, $\bf \Phi_1$ and $\bf a$ into
\begin{equation}
{\bf{u}} = \left[ {\begin{array}{*{20}{c}}
{{{\bf{u}}_{\bf{A}}}}\\
{{{\bf{u}}_{\bf{B}}}}
\end{array}} \right], {\kern 2pt}
{\bf G} = \left[ {\begin{array}{*{20}{c}}
{{{\bf G} _{\bf{A}}}}\\
{{{\bf G} _{\bf{B}}}}
\end{array}} \right], {\kern 2pt}
{\bf{a}} = \left[ {\begin{array}{*{20}{c}}
{{{\bf{a}}_{\bf{A}}}}\\
{{{\bf{a}}_{\bf{B}}}}
\end{array}} \right],
\label{eq_85}
\end{equation}
where ${\bf u_A} \in {\cal C}^{N \times 1}$ consists of non-negative dual variables and can be expressed as
\begin{equation}
{\bf u_A}=\frac{1}{2}{\bf G_A}{{\bf{q}}} + \frac{{\bf{a_A}}}{c}.
\label{eq_86}
\end{equation}
Based on the above transformation and the iterative algorithm proposed in \cite{r45}, we propose an iterative algorithm for the generic QP formulation in ${\cal P}_{13}$, given in Algorithm 1, where most of the procedure follows \cite{r45}. The proposed algorithm is applicable to both scenarios $K \le N_t$ and $K > N_t$, by constructing the coefficient matrix $\bf Q$ as ${\bf Q}={\bf \tilde V}^{-1}$ in \eqref{eq_53} and as ${\bf Q}={\bf FDY}^{-1}{\bf D}^T{\bf F}^T$  in \eqref{eq_69}, respectively.

\newcounter{mytempeqncnt2}
\begin{figure*}
\normalsize
\setcounter{mytempeqncnt2}{\value{equation}}
\setcounter{equation}{93}

\begin{equation}
{\bf{W}} = \left\{ {\begin{array}{*{20}{c}}
{\frac{1}{K} \cdot {{\bf{H}}^H}{{\left( {{\bf{H}}{{\bf{H}}^H}} \right)}^{ - 1}}{\bf{U}} \cdot diag\left\{ {\sqrt {\frac{{{p_0}}}{{{{\bf{1}}^T}{\bf{\tilde V1}}}}}  \cdot {{\bf{F}}^{ - 1}}{\bf{1}}} \right\}{{\bf{s_E}}}{{{\bf{\hat s}}}^T}}, {\kern 3pt} {\rm{if}} {\kern 3pt} d \ge 0 \\
{\frac{1}{K} \cdot {{\bf{H}}^H}{{\left( {{\bf{H}}{{\bf{H}}^H}} \right)}^{ - 1}}{\bf{U}} \cdot diag\left\{ {\sqrt {\frac{{{p_0}{e^2}}}{{\left( {d \cdot {\bf{g}}_k^T + e \cdot {{\bf{1}}^T}{\bf{\tilde V}}} \right)\left( {d \cdot {{{\bf{\tilde V}}}^{ - 1}}{{\bf{g}}_k} + e \cdot {\bf{1}}} \right)}}} {{\bf{F}}^{ - 1}}\left( {d \cdot {{{\bf{\tilde V}}}^{ - 1}}{{\bf{g}}_k} + e \cdot {\bf{1}}} \right)} \right\}{{\bf{s_E}}}{{{\bf{\hat s}}}^T}}, {\kern 3pt} {\rm{if}} {\kern 3pt} d < 0
\end{array}} \right.
\label{eq_89}
\end{equation}

\setcounter{equation}{\value{mytempeqncnt2}}
\hrulefill
\vspace*{4pt}
\end{figure*}

\begin{algorithm}
  \caption{Proposed Algorithm for a Generic QP Optimization ${\cal P}_{13}$}
  \begin{algorithmic}
    \State ${\bf input:}$ ${\bf Q}$
    \State ${\bf output:}$ ${\bf u}^*$
    \State ${\bf i}=[{\kern 2pt}]$, $I={\rm{length}}\left( {\bf{i}} \right)$, ${\bf N}=[ 1 ]$, $n=1$, and ${\rm{Iter}}=0$;
    \State Calculate ${\bf a}={\bf Q}^{-1}{\bf 1}$, $c={\bf 1}^T{\bf a}$; 
    \State Obtain $\bf G$ in \eqref{eq_84}, ${\bf u}=\frac{\bf a}{c}$ and ${\bf u_A}=\frac{\bf a_A}{c}$;
    \If {${\rm{min}}\left( {{{\bf{u}}_{\bf{A}}}} \right) < 0$}
    \State Set ${\bf u}^*={\bf u}$;
    \Else
    \While {$\min \left( {\bf{u_A}} \right) < 0$ and ${\rm{Iter}}<{\rm{Iter}}_{\rm max}$}
    \State ${\bf{d}} = {\rm{sort}}\left( {{{\bf{u_A}}}} \right)$;
    \State find $\mu_k=d_{n}$; Stack ${\bf{N}} = \left[ {\begin{array}{*{20}{c}}{\bf{N}}&1 \end{array}} \right]$, ${\bf{i}} = \left[ {\begin{array}{*{20}{c}}
{\bf{i}}&k
\end{array}} \right]$;
    \State Update $I$;
    \State Construct ${\bf{Z}} = \left[ {\begin{array}{*{20}{c}}
{{\bf{G}}\left( {{i_1},{i_1}} \right)}& \cdots &{{\bf{G}}\left( {{i_1},{i_I}} \right)}\\
 \vdots & \ddots & \vdots \\
{{\bf{G}}\left( {{i_I},{i_1}} \right)}& \cdots &{{\bf{G}}\left( {{i_I},{i_I}} \right)}
\end{array}} \right]$;
    \State Set ${{\bf{\tilde a}}_{\bf{A}}} = {\left[ {{{\bf{a}}_{\bf{A}}}\left( {{i_1}} \right),{{\bf{a}}_{\bf{A}}}\left( {{i_2}} \right), \cdots ,{{\bf{a}}_{\bf{A}}}\left( {{i_I}} \right)} \right]^T}$;
    \State Calculate ${\bf{\tilde q}} =  - \frac{2}{c} \cdot {{\bf{Z}}^{ - 1}}{{\bf{\tilde a}}_{\bf{A}}}$;
    \If {$\min \left( {{\bf{\tilde q}}} \right) \ge 0$}
    \State Update ${\bf{u}}$ and $\bf u_A$;
    \State $n=1$;
    \Else
    \State find ${q_k} = \min \left( {{\bf{\tilde q}}} \right)$ and $i_m=k$;
    \State Set ${\bf{i}} = {\bf{i}}\left( {1:m} \right)$ and ${\bf{N}} = {\bf{N}}\left( {1:m} \right)$; 
    \State Update $I$, $\bf Z$, $\bf \tilde a$, $\bf \tilde q$, $\bf u$, and ${\bf{u_A}}$;
    \State Update ${\bf{N}}\left( {m} \right) \leftarrow {\bf{N}}\left( {m} \right) + 1$;
    \State Update $n = {\bf{N}}\left( {m} \right)$;
    \EndIf
    \State ${\rm{Iter}} \leftarrow {\rm{Iter}}+1$;
    \EndWhile
    \State Obtain ${\bf u}^*={\bf u}$;
    \EndIf
   \end{algorithmic}
\end{algorithm}

\subsection{The Sub-Optimal Closed-Form Precoder}
While the algorithm proposed above includes a closed-form expression within each iteration, it is still an iterative algorithm in nature. Therefore, we further propose a closed-form precoder that achieves sub-optimal performance, for both scenarios considered in this paper. The proposition of this sub-optimal precoder is based on the observation that there is at most only one entry in ${\bf q_E}$ that is non-zero for some of the channel realizations.

It has been demonstrated in {\bf Observation 3} that the CI precoder will reduce to the ZF precoder when ${\bf q_E}$ is a zero vector. Therefore in this section, we focus on the scenario where only one entry in ${\bf q_E}$ is non-zero. This corresponds to the case where only one entry in $\bf a_A$ is negative, which violates the constraints in ${\cal P}_{13}$ if ${\bf q_E}$ is a zero vector. Without loss of generality, we assume $d=\min \left( {{{\bf{a}}_{\bf{A}}}} \right) < 0$ and $k$ is the corresponding index, i.e., $d={{\bf{a}}_{\bf{A}}}\left( k \right) = \min \left( {{{\bf{a}}_{\bf{A}}}} \right)$. We can then obtain $\mu_k = 0$ and $q_k \ne 0$ based on ${\cal P}_{14}$. Subsequently, based on \eqref{eq_86}, we can further express
\begin{equation}
\begin{aligned}
&{\kern 3pt} {\mu _k} = {{\bf{u}}_{\bf{A}}}\left( k \right) = \frac{1}{2}{{\bf{G}}_{\bf{A}}}\left( {k,k} \right){q_k} + \frac{{{{\bf{a}}_{\bf{A}}}\left( k \right)}}{c} = 0\\
\Rightarrow & {\kern 3pt} {q_k} =  - \frac{{2 {{\bf{a}}_{\bf{A}}}\left( k \right)}}{{c {{\bf{G}}_{\bf{A}}}\left( {k,k} \right)}},
\end{aligned}
\label{eq_87}
\end{equation}
which is based on the assumption that only one entry in $\bf q$ is non-zero, otherwise the complementary slackness condition will not be satisfied. Accordingly, the resulting sub-optimal solution for $\bf u$ can be obtained based on \eqref{eq_84}, given by
\begin{equation}
\begin{aligned}
{\bf{u}} &= \frac{1}{2}{{\bf{g}}_k} \cdot \left[ { - \frac{{2{{\bf{a}}_{\bf{A}}}\left( k \right)}}{{c{{\bf{G}}_{\bf{A}}}\left( {k,k} \right)}}} \right] + \frac{{\bf{a}}}{c}\\
& =  - \frac{{{{\bf{a}}_{\bf{A}}}\left( k \right) \cdot {{\bf{g}}_k} + {{\bf{G}}_{\bf{A}}}\left( {k,k} \right) \cdot {\bf{a}}}}{{c{{\bf{G}}_{\bf{A}}}\left( {k,k} \right)}}\\
& =  - \frac{{d \cdot {{\bf{g}}_k} + e \cdot {\bf{a}}}}{{ce}}\\
\end{aligned}
\label{eq_88}
\end{equation}
where we decompose ${{\bf{G}}_{\bf{A}}} = {\left[ {{\bf{g}}_1,{\bf{g}}_2, \cdots ,{\bf{g}}_N} \right]}$, ${\bf g}_k$ is the $k$-th column of $\bf G_A$, and $e={{{\bf{G}}_{\bf{A}}}\left( {k,k} \right)}$. Based on the obtained sub-optimal $\bf u$ in \eqref{eq_88} and by substituting ${\bf Q}={\bf \tilde V}^{-1}$, we are able to express the sub-optimal closed-form CI precoder for the case of $K \le N_t$ in \eqref{eq_89} at the top of this page. The sub-optimal closed-form CI precoder for the case of $K > N_t$ can be obtained in a similar form and is therefore omitted for brevity.

\section{Numerical Results}
In this section, the numerical results of the proposed schemes are presented and compared with traditional CI precoding using Monte Carlo simulations. In each plot, we assume the total transmit power is $p_0=1$, and the transmit SNR per antenna is thus $\rho = {1 \mathord{\left/ {\vphantom {1 {{\sigma ^2}}}} \right. \kern-\nulldelimiterspace} {{\sigma ^2}}}$. For the case of $K \le N_t$, we compare our proposed iterative schemes with traditional ZF precoding, RZF precoding, and optimization-based CI precoding approach. For the case of $K > N_t$, we compare our proposed iterative algorithm with RZF precoding and the optimization-based CI precoding method. Both 16QAM and 64QAM modulations are considered in the numerical results.

The following abbreviations are used throughout this section:

\begin{enumerate}
\item `ZF': traditional ZF scheme with symbol-level power normalization for $K \le N_t$ \cite{r5};
\item `RZF': traditional RZF scheme with symbol-level power normalization for both $K \le N_t$ and $K > N_t$ \cite{r6};
\item `CI-OPT': traditional optimization-based CI precoding based on ${\cal P}_{3}$;
\item `CI-Iterative': iterative CI precoding scheme based on Algorithm 1;
\item `CI-CF': sub-optimal closed-form CI precoder introduced in Section V-A.
\end{enumerate}

\begin{figure}[!b]
\centering
\includegraphics[scale=0.4]{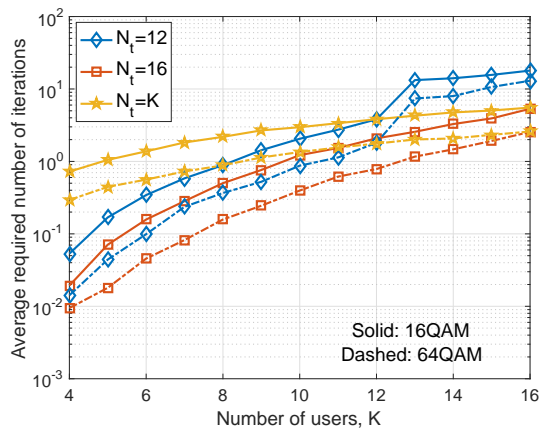}
\caption{Average required number of iterations v.s. number of users $K$, $N_t=12$, $N_t=16$ and $N_t=K$, 2000 trials}
\end{figure} 

Before we present the detailed bit error rate (BER) results, in Fig. 3 we first show the average number of iterations required for the proposed algorithm to achieve the optimality with an increasing number of users, where we consider the case of `$N_t=12$' and `$N_t=16$', as well as the case of `$N_t=K$'. Generally, we observe that the required number of iterations increases with the number of users, since there is a higher possibility that more entries in $\bf a_A$ are negative. Comparing the result of `$N_t=16$' with `$N_t=K$', we observe that the required number of iterations becomes smaller when $K<N_t$. This is because the channel becomes closer to orthogonal when $\left( {N_t - K} \right)$ is larger, in which case it is very likely that ZF precoding is the optimal solution and the resulting required number of iterations is 0. Moreover, for the case of `$N_t=12$', we observe a significant increase in the number of iterations when $K \ge 13$, which is because the scenario is shifted from `$K \le N_t$' to `$K>N_t$', where the formulation $\bf Q$ in Algorithm 1 follows \eqref{eq_69} instead of \eqref{eq_53}. Comparing the results of 16QAM with 64QAM, it is also observed that the required number of iterations for 64QAM is smaller than that for 16QAM, as the optimal CI precoding is more likely to be a ZF precoder for 64QAM. In the following, we present the BER results for the scenarios of both $K \le N_t$ and $K > N_t$. For the case of $K \le N_t$, we focus on the symmetric case $K=N_t$ which is the most challenging.

\subsection{$K \le N_t$}

\begin{figure}[!t]
\centering
\includegraphics[scale=0.4]{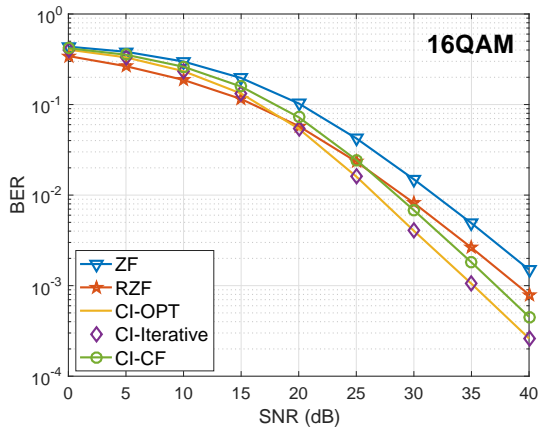}
\caption{Uncoded BER v.s. SNR, 16QAM, $N_t=K=8$}
\end{figure} 

Fig. 4 presents the BER performance for different precoding schemes with 16QAM modulation, where $K=N_t=8$. As can be observed, the interference-exploitation precoding achieves an improved performance over ZF precoding for all SNRs, while also outperforming RZF precoding at high SNR. For high SNRs, we observe a SNR gain of more than 6dB over ZF and 4dB over RZF. Moreover, it is shown that the CI precoding based on the iterative algorithm `CI-Iterative' achieves exactly the same performance as the optimization-based CI precoding `CI-OPT', which validates its optimality in obtaining the precoding matrix. While the sub-optimal closed-form precoder `CI-CF' is inferior to optimal CI precoding, we observe that it also outperforms both ZF precoding and RZF precoding when the SNR is high.

\begin{figure}[!t]
\centering
\includegraphics[scale=0.4]{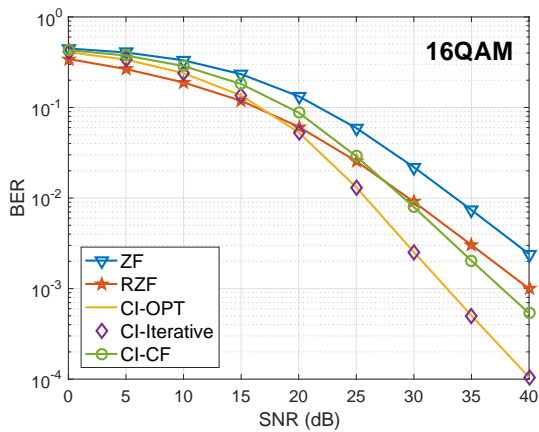}
\caption{Uncoded BER v.s. SNR, 16QAM, $N_t=K=12$}
\end{figure} 

\begin{figure}[!t]
\centering
\includegraphics[scale=0.4]{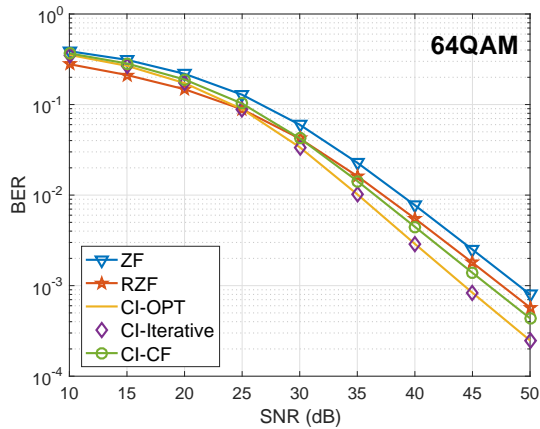}
\caption{Uncoded BER v.s. SNR, 64QAM, $N_t=K=12$}
\end{figure} 

In Fig. 5, we compare the BER of different precoding approaches for 16QAM when $K=N_t=12$, and a similar BER trend compared to the case of $K=N_t=8$ in Fig. 4 is observed. The optimal CI precoding scheme achieves the best performance and significantly outperform other precoding methods in the high SNR regime, where the SNR gain is as large as 7dB. Compared with Fig. 4, we observe that the performance gains of CI precoding over traditional ZF-based precoding schemes are more significant when the number of users and antennas increases.

We further consider a higher-order 64QAM modulation and present the corresponding BER result in Fig. 6. Similar to the case of 16QAM, both the optimal CI precoding and the sub-optimal closed-form CI precoding methods achieve an improved performance over ZF precoding. More importantly, in contrast to claims that CI precoding may not be promising for higher-order QAM modulations where only the outer constellation points can exploit CI, we show that the performance gains of CI precoding over ZF can be as large as 5dB, since CI precoding not only relaxes the optimization area for the outer constellation points, but also reduces the noise amplification factor for the precoder. Both of the above contribute to the performance improvements over ZF precoding.

\subsection{$K > N_t$}

\begin{figure}[!t]
\centering
\includegraphics[scale=0.4]{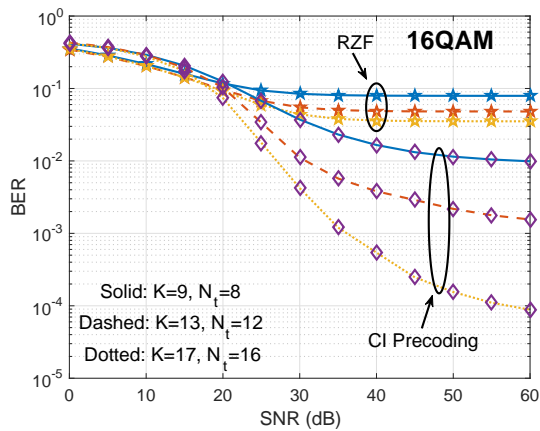}
\caption{Uncoded BER v.s. SNR, 16QAM, $K > N_t$}
\end{figure} 

In this section, we consider the scenario of $K > N_t$. Fig. 7 depicts the BER result for (1) $K=9$, $N_t=8$, (2) $K=13$, $N_t=12$, and (3) $K=17$, $N_t=16$ for 16QAM modulation. ZF precoding is not applicable in these cases, and therefore we compare with RZF precoding. When CI precoding does not return a valid solution, as discussed in Section IV-A, RZF precoding is employed instead. For all of the three considered scenarios, we observe that at high SNRs the average BER performance for CI precoding achieves a significant performance gain over traditional RZF precoding, since there is a high probability of a feasible solution for CI precoding when $K-N_t=1$, as illustrated later in Fig. 9. Moreover, we observe that the performance gains further increase with an increase in the number of transmit antennas.

\begin{figure}[!t]
\centering
\includegraphics[scale=0.4]{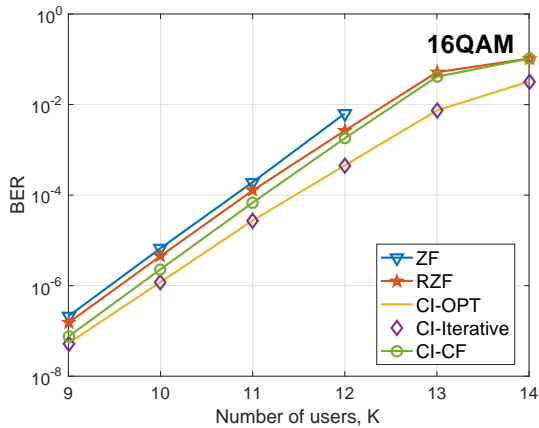}
\caption{Uncoded BER v.s. number of users $K$, 16QAM, $N_t=12$, SNR=35dB}
\end{figure} 

We present the BER result with respect to an increasing number of users in Fig. 8 for 16QAM, where $N_t=12$ and SNR=35dB. ZF precoding does not apply for the case of  $K>N_t$ and therefore does not have a BER result when $K>12$. For both the case of $K \le N_t$ and $K > N_t$, we observe that CI precoding achieves the best BER performance, and the performance gains over ZF-based precoding are more significant when $K$ increases, which is due to the fact that the optimal CI precoding is more likely to be ZF precoding when $K$ is small.

In Fig. 9, we present the average feasibility probability for the case of $K > N_t$ with an increasing number of users, where the performance of 16QAM and 64QAM modulations are denoted as solid and dashed lines, respectively. Generally, we observe that a larger number of antennas at the BS leads to a higher probability of supporting more users than the number of antennas at the BS. For example, for 16QAM modulation, the BS can support 3 more users when $N_t=12$ with more than a 70\% feasibility probability, compared to the $N_t=6$ case where the BS can only support at most only 1 additional user with the same feasibility probability. Specifically, we observe that the feasibility probability for the case of $K=13$ and $N_t=12$ is very close to 100\% for 16QAM. A similar trend is also observed for 64QAM modulation, although the supported number of users is smaller compared to 16QAM for the same feasibility probability.

\begin{figure}[!t]
\centering
\includegraphics[scale=0.4]{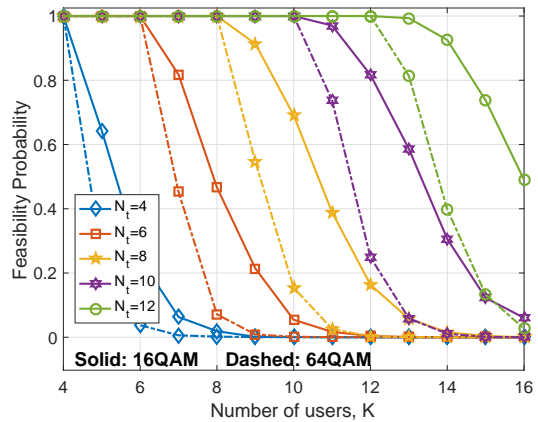}
\caption{Feasibility probability v.s. number of users $K$}
\end{figure} 

\section{Conclusion}
In this paper, interference-exploitation precoding for multi-level modulations is studied. By analyzing the optimization problems and their corresponding Lagrangian and KKT conditions, we mathematically show that interference-exploitation precoding is equivalent to a QP optimization, based on which we derive the optimal precoding matrix as a function of the variables for the QP optimization as well as a sub-optimal closed-form precoder. Numerical results show that the optimal precoding matrix can be efficiently obtained by the proposed iterative algorithm, and the superior performance improvement of interference-exploitation precoding over traditional precoding schemes is also observed for multi-level modulations.

\section*{Acknowledgment}
The authors would like to thank Dr. Fan Liu for his valuable suggestions.

\ifCLASSOPTIONcaptionsoff
  \newpage
\fi

\bibliographystyle{IEEEtran}
\bibliography{refs.bib}

\end{document}